\documentstyle[11pt,amssym,aaspp4,flushrt]{article}

\lefthead{GEROYANNIS}
\righthead{SPIN-DOWN OF THE PRIMARY IN AE AQUARII}

\begin{document}
\title{THE SPIN-DOWN PROBLEM\\
       IN THE WHITE DWARF OF AE AQUARII:\\
       NUMERICAL STUDY OF A TURN-OVER SCENARIO}
\author{V. S. Geroyannis}
\affil{Astronomy Laboratory, Deparment of Physics, University of Patras,\\
	GR--26110 PATRAS, GREECE\\ 
	Author's e-mail: vgeroyan@physics.upatras.gr}

\begin{abstract}
The white dwarf (i.e., the primary) in the AE Aqr binary star system is observed to spin down at a steady time rate $\dot{P}_{now} = 5.64 \times 10^{-14} \, \rm s \, s^{-1}$, while at the same time its UV $(\frak{L}_{UV})$ and X-ray $(\frak{L}_X)$ accretion luminosities seem to remain almost unchanged. This is a contradiction, however, since the classically estimated spin-down power due to decrease of rotational kinetic energy $T$, $\dot{T} \sim - 10^{34} \, \rm erg \, s^{-1}$, exceeds the accretion luminosity of the primary by a factor $\gtrsim 10^2$ and, as a dominating power, should lead either to observable luminosity changes or to other detectable effects. This so-called ``spin-down problem'' can be relaxed under the assumption that the primary is now in a phase of rapidly decreasing its differential rotation under constant angular momentum, undergoing a nonaxisymmetric transition --- i.e., turning over its magnetic symmetry axis with respect to its angular momentum axis and eventually becoming a perpendicular rotator --- from a ``differential rotation state'' (DRS) to a ``rigid rotation state'' (RRS) on a relatively short ``nonaxisymmetric DRS-to-RRS transition timescale''. If so, then the estimated spin-down power, $- \dot{T} \lesssim 4 \times 10^{32} \, \rm erg \, s^{-1}$, becomes comparable to the observed luminosities and the spin-down problem is drastically simplified. In this paper, we present a detailed numerical study of such a ``turn-over scenario'', which study mainly points out the fact that an observed large spin-down time rate does not always imply a large spin-down power.        
\end{abstract}
\keywords{stars: individual (AE Aqr) --- stars: magnetic fields --- stars: rotation and spin-down --- stars: white dwarfs}

\section{Introduction}
AE Aqr has been studied extensively in the last years (see, e.g., \cite{whgo93}; \cite{whok93}; \cite[hereafter dJ\&94]{jmor94}; \cite[hereafter E\&94]{ehrz94}; \cite[hereafter M\&94]{mjrn94}; \cite[hereafter P94]{patt94}; \cite[hereafter I95]{ikhs95}; \cite[hereafter WHG95]{whgo95}; \cite[hereafter C\&96]{cmmh96}; \cite[hereafter EH96]{ehor96}; \cite[hereafter I97]{ikhs97}; \cite[hereafter K\&97]{kfah97}; \cite[hereafter WKH97]{wkho97}; \cite[hereafter I98]{ikhs98}; \cite{whgo98}; \cite{cdag99}; \cite[hereafter I99]{ikhs99}). 

AE Aqr has been classified (see, e.g., P94, \S \, 2.4.2) as member of the DQ Her or Intermediate Polar (IP) cataclysmic variables (CVs). Such CVs are binary star systems in which a rotating magnetic white dwarf (playing the role of ``primary'') and a late-type main-sequence star (playing the role of ``secondary'') are adjusted in a highly confined orbit (i.e., orbit with orbital separation less than a few solar radii). The secondary extends out to its critical Roche equipotential surface and small motions of thermal origin in its atmosphere lead to gas flow (through the inner Lagrangian point of the binary) into the potential well of the primary. This gas forms an accretion disk, which is an efficient machine at converting gravitational energy to radiation and heat, and at redistributing angular momentum. The accretion disk is disrupted at the Alfv\'{e}n radius and, eventually, the accretion flow is channelled towards the poles of the primary through magnetic columns (P94, \S \, 2.1).

However, AE Aqr is characterized by some unusual properties. We mention here its violent flaring activity, with recurrence timescale $\sim 1 \rm \, h$, from the ultraviolet (UV) (E\&94, \S \S \, 2--4) to the radio frequencies (\cite[\S \, 2]{bdch88}). It is also remarkable that its spin period, $P_{now}=33.08 \rm \, s$ (the shortest one among IPs), has been detected and verified over a wide range of frequencies, from the optical (\cite[\S \S  \, 2, 6]{patt79}) to the $\gamma$-rays (\cite[\S \, 2]{mrjb92}, M\&94, \S \S \, 2--3, 6). Further unusual properties of AE Aqr are discussed by, e.g., I95 (\S \, 1), WHG95 (\S \, 1), C\&96 (\S \, 1), \cite[(\S \S \, 2, 4--5)]{bgru97}, I97 (\S \, 1), K\&97 (\S \, 1), and WKH97 (\S \, 1). 

Concerning the subject of the present investigation, emphasis is given on the puzzling rapid spin-down time rate, $\dot{P}_{now} = 5.64 \times 10^{-14} \, \rm s \, s^{-1}$, reported by dJ\&94 ( \S \S \, 6--7, Table 4). The implications of such a spin-down have been discussed by several investigators (see, e.g., dJ\&94, \S \, 8; WHG95, \S \, 1; \cite[hereafter B\&96, \S \, 4]{bibs96}; I97, \S \S \, 1--3; I98, \S \S \, 2--5). Especially in I97 (\S \, 1.2), the so-called ``spin-down problem'' is defined clearly. In particular, when calculated classically, the observed spin-down time rate $\dot{P}_{now}$ implies a spin-down power of the white dwarf (I97, eq. [1])
\begin{equation} \label{eq:Tdot}
\dot{T} \simeq - 6 \times 10^{33} I_{50} \dot{P} P^{-3} \, \rm erg \, s^{-1}
        \sim - 10^{34}\, \rm erg \, s^{-1},
\end{equation}
where $I_{50}$ is the moment of inertia of the white dwarf in units of $10^{50} \, \rm g \, cm^2$, and $\dot{P}$ and $P$ are expressed in units of $\dot{P}_{now}$ and $P_{now}$, respectively. Comparing $\dot{T}$ with the observed luminosities (see, e.g., I97, Table 1), we realize that it exceeds the UV $(\frak{L}_{UV})$ and X-ray $(\frak{L}_X)$ luminosities of AE Aqr by a factor  $\gtrsim 10^2$. This is in contradiction with the basic postulate of the IP model concerning the accretion nature of the primary's radiation (for details on this issue, see I95, \S \, 2.3), and exactly this contradiction has been termed as ``spin-down problem''.

In the following two sections, we shall discuss some scenarios proposed for solving this problem; we shall also describe the scenario considered in the present investigation.

\section{Dealing with the spin-down problem}
It is of interest, at this point, to mention some models concerning the magnetic, radiating, and accreting properties of AE Aqr; although some of these models are not directly connected to the spin-down problem, they help us to have a better understanding of the whole matter. In I95 (\S \S \, 2--3, Table 1), the state of the primary (under the basic assumption that it is a white dwarf) is taken to be simultaneously as ejector, propeller, and accretor. Then the spin-down power may be lost in the form of relativistic stellar wind ejected from the surface of the white dwarf (ejector) or transmitted to the accretion flow near the Alfv\'{e}n surface (combination of accretor and propeller; see also EH96, \S \, 4.3). 

In K\&97 (\S \, 8), it is concluded that if the rapidly rotating white dwarf in AE Aqr has a surface magnetic field of $\sim 100 \, \rm T$, then magnetic pumping is a serious candidate for the particle acceleration needed to explain the observed radio outbursts. Magnetic pumping (K\&97, \S \, 4 and references therein) is one of a number of stochastic particle acceleration mechanisms, known also as Fermi-type mechanisms, whose essential property is the combination of a particle interaction with a spatial or temporal change in the magnetic field --- which causes the particle energy to change --- with a randomizing process (see also EH96, \S \, 5). 

WKH97 (\S \S \, 2--3, 6) discuss in detail the so-called ``diamagnetic blob model'' (for some related topics of interest, see also \cite{klas91}, \cite{wkin92}, \cite{king93}, \cite{wkin95}). According to the authors, the discrepancy between the large spin-down power of the white dwarf and the observed luminosity of the system is explained by the kinetic energy carried away by the ejecta. This model does also predict a magnetic dissipation time rate $\sim 10^{33} \, \rm erg \, s^{-1}$. There are some processes that may be triggered by such dissipation, but they could account for only $\lesssim 1$ percent of the dissipated energy. A significant fraction of this energy may go into particle acceleration and production of $\gamma$-rays.

In I97, a neutron star model is proposed for the primary in AE Aqr. The author concludes that, in the frame of the wind-fed accretor model, there are no particular problems with the interpretation of the observed spin-down time rate of the primary (remember that the moment of inertia of a neutron star is $\sim 5$ orders of magnitude less than that of a white dwarf). However, this scenario reveals a very intriguing problem (I97, \S \, 4): if the primary in AE Aqr is indeed a neutron star, why its appearance in the optical UV is very likely to that of an accreting white dwarf?

In a subsequent investigation, Ikhsanov (I98, \S 3) recognizes that, in spite of some progress, the origin of the observed spin-down of the primary in AE Aqr cannot be regarded as well understood so far; in this situation the investigation of alternative possibilities (if they exist) remains important. In this view, the author (I98, \S \S \, 4--5) discusses a pulsar-like white dwarf model in a binary system. In particular, he explores the possibility to solve the spin-down problem of AE Aqr within the ``ejector approach'', assuming that the major part of the rotational kinetic energy of the white dwarf is spent in the generation of low frequency magnetodipole waves and particle acceleration. As the author discusses, the main consequence of this approach is a relatively strong magnetic field of the white dwarf, which makes this model to differ radically from previous ones. The estimated magnetic moment ($\sim 10^{34} \, \rm G \, cm^{3}$) is $\sim 2$ orders of magnitude greater than those of previous models; and the question arising concerns the consistency of such an estimate with corresponding observations (see I98, \S \, 5 for interesting details and references therein; especially, \cite{crop86}, \cite{ssbl92}, B\&96).

Recently, Ikhsanov (I99) describes a scenario according to which ``magnetic amplification'' can be achieved by the action of differential rotation, appearing, in turn, as consequence of gravitational waves (and resulting angular momentum loss); if so, then the magnetic field inside the star winds up, on a timescale of $\sim 1 \, \rm month$, up to $\sim 10^8 \div 10^9 \, \rm G$ and the star spins down predominantly due to magneto-dipole waves and particle acceleration. Furthermore, in the framework of the pulsar-like white dwarf model (as in I98, I99), \cite{ikhs00} suggests the existence of a dead disk around the white dwarf magnetosphere. The main feature that makes the suggested model different from those previously discussed, is the hot boundary layer at the inner disk radius. Due to this feature, the model sheds new light to some puzzling properties of AE Aqr.

\cite{mdja00}, in order to explain the multiwavelength emission from AE Aqr, propose a model on the basis of the assumption that the compact component of this binary system consists of a white dwarf with a surface field length of $\sim 1 \, \rm MG$, rotating through a clumpy ring near the circularization radius. After a detailed analysis, they conclude that the whole reservoir of the spin-down power has gone into the production of very high-energy particles and $\gamma$-ray emission.

Finally, \cite{chyi00} point out that the electromagnetic dipole radiation with surface magnetic field $\lesssim 5 \times 10^6 \, \rm G$ and the propeller action with accretion time rate $\lesssim 10^{17} \, \rm g \, s^{-1}$ are unable to account for the observed spin-down torque. So, the authors propose that the spin-down power does not transform into the observed electromagnetic radiation and they consider the gravitational radiation emission as an alternative spin-down mechanism. The authors remark that this mechanism could be an attractive one since the resulting spin-down power does not need to go into the observed electromagnetic radiation, which effectively avoids the long standing question of nondetection of the observed large spin-down power.

\section{The turn-over scenario}
In the present paper, we try to resolve the spin-down problem in terms of intrinsic properties (like angular momentum, moments of inertia, differential rotation, and magnetic field) of the primary in AE Aqr. In particular, we propose an axisymmetric, differentially rotating, magnetic, white-dwarf model with its surface magnetic field causing a slow ``secular spin-down'' (to be distinguished from the fast spin-down observed now) and, accordingly, a slow ``secular angular momentum loss''. This model participates simultaneously in the accretional activities of the binary system AE Aqr; and, apparently, its secular activities are screened by its accretional activities as long as the former remain weak in comparison with the latter. Under certain circumstances, however, such secular activities can cause certain quite intense phenomena. In this view, we shall try to document the aspect that the fast spin-down of the primary in AE Aqr, observed now, belongs to such a category of phenomena. In particular, we shall give numerical results showing that such fast spin-down may be the result of a nonaxisymmetric transition from an initial axisymmetric differential rotation state to a terminal nonaxisymmetric rigid rotation state, due to turn-over of the magnetic symmetry axis with respect to the invariant angular momentum axis (the latter taken to be the $z$ axis). Our results mainly show that a fast spin-down does not always imply a large spin-down power; instead, under certain circumstances as this of turning over, the spin-down power corresponding to a large spin-down time rate (as that observed in AE Aqr) is $\sim 2$ orders of magnitude less than that estimated classically.      

We shall use hereafter definitions and symbols identical to those in \cite{gpap99}. 

According to the so-called ``turn-over scenario'' (hereafter TOV scenario), the model undergoes an axisymmetric ``early evolutionary phase'', during which the combined effect of rotation and (toroidal plus poloidal) magnetic field derives configurations with moment of inertia along the rotation axis $I_3$ (coinciding with the symmetry axis), $I_{33}$, greater than the moment(s) of inertia along the other two principal axes $I_1$ and $I_2$, $I_{11} = I_{22}$; i.e., $I_{33} > I_{11}$. At some particular time, however, which signals the beginning of a ``late evolutionary phase'', secular angular momentum loss yields configurations with $I_{33}(L_x) = I_{11}(L_x)$ (relation valid for a particular angular momentum $L_x$) and, after that particular time, configurations with $I_{33}(L) < I_{11}(L)$ when $L < L_x$. Thus, a ``dynamical asymmetry'' is established in the sense that the moment of inertia along the rotation axis becomes less than the moment(s) of inertia along the other two principal axes. This inequality is due to the action of toroidal magnetic field, which tends to derive prolate configurations; while, on the other hand, rotation and poloidal magnetic field tend to derive oblate configurations. 

Eventually, secular decrease of rotation due to surface magnetic field gives rise to dynamical asymmetry, $I_{11} > I_{33}$. However, dynamically asymmetric configurations tend to turn over spontaneously, so that to subsequently rotate along axis with moment of inertia greater than $I_{33}$. They become, therefore, ``oblique rotators''; the well-known ``perpendicular rotator'' is an oblique rotator with ``turn-over angle'' (or, equivalently, ``obliquity angle'': the angle of the magnetic symmetry axis with respect to the invariant angular momentum axis), $\chi$, equal to $90 \arcdeg$. The reasons causing turn-over of dynamically asymmetric configurations have been studied by several investigators (see, especially, \cite[hereafter MT72]{mtak72}); details on this issue are given in \S \, 6. 

Within the framework of TOV scenario, we shall study the case in which the terminal nonaxisymmetric rigidly rotating model becomes perpendicular rotator on a ``turn-over timescale'', $t_{TOV}$, comparable with the ``period timescale'', $t_{P} = P_{now}/\dot{P}_{now}$; i.e., $t_{TOV} \sim t_{P}$. Although any intermediate oblique model with $\chi < 90 \arcdeg$ undergoes a complicated rotation along axis passing through the center of mass but not coinciding with one of its principal axes (MT72, \S\S \, 2--3), the terminal model rotates along its $I_1$ principal axis, which axis is now coinciding with the invariant angular momentum axis taken, in turn, to be the $z$ axis. Apparently, its angular momentum vector remains equal to that of the starting model (MT72, \S \S \, 1--2), while its angular velocity vector obtains same direction with that of the starting model (coinciding with the angular momentum direction) but different magnitude. 

During the turn-over phase, the turn-over angle increases spontaneously up to $90 \arcdeg$ on timescale $t_{TOV}$, since the rotational kinetic energy of the model decreases from a higher level with turn-over angle $\chi \gtrsim 0 \arcdeg$ to a lower level with turn-over angle $\chi \simeq 90 \arcdeg$. At this level, the model reaches the state of least energy consistent with its prescribed angular momentum and magnetic field. Accordingly, the excess energy due to the differential rotation field of motions, defined by the angular velocity component $\Omega_3$ along the (spontaneously turning over) $I_3$ principal axis (coinciding, in turn, with the magnetic symmetry axis), is dissipated down to zero under the action of turbulent viscosity in the convective regions of the model (MT72, \S \S \, 3--4). Thus, differential rotation along the $I_3$ principal axis decreases down to zero and all of its effects disappear.

Furthermore, it seems difficult for an oblique model to sustain differential rotation along its $I_1$ principal axis mainly due to the destructive action of the poloidal magnetic field (see, e.g., \cite[hereafter MMT90, \S\S \, 1--3]{mmta90}; also, \cite[hereafter M92, \S\S \, 1, 5-6]{moss92}). In simple words, in a model with magnetic field symmetry axis inclined with respect to the rotation axis, i.e., $\chi > 0 \arcdeg$, nonuniform rotation in the electrically conducting stellar material shears the field and generates magnetic torques. As a result, there is a competition between the efforts of the magnetic stresses to remove rotational nonuniformities, and those of the rotational velocities to bury and destroy magnetic flux. If, then, the magnetic field and the electrical conductivity have appropriate values (issues to be discussed in \S \, 7; see, e.g., MMT90, \S \, 4; also, M92, \S \, 6), then the magnetic field prevails and removes all the nonuniformities of rotation. Consequently, the terminal model rotates rigidly along its $I_1$ principal axis with angular velocity $\Omega_1$. However, differential rotation behaves like a ``rotational kinetic energy capacitor'' and its decrease down to zero discharges all this excess energy (since the rigidly rotating perpendicular rotator has a much lower ``rotational kinetic energy capacity''). Hence, differential rotation decrease leads to releasing the excess rotational energy on a ``differential rotation decrease timescale'', $t_{DRD} \sim t_{TOV}$,. Our task here is to verify that the corresponding energy release time rate is $\sim 10^2$ times less than that estimated by equation (\ref{eq:Tdot}). 

In detail, we assume that the model of TOV scenario undergoes a nonaxisymmetric transition from its axisymmetric DRS to its nonaxisymmetric perpendicular RRS. The starting model has mass $M_{DR} = 0.89 M_{\sun}$, angular momentum $L_{DR} = L_{xx}$ such that $L(P_{now}) < L_{xx} < L_x$, rotational kinetic energy $T_{DR} = T_{xx}$, spin period $P_{DR} = P_{xx} < P_{now}$, magnetic perturbation parameter $h_{DR}$, and magnetic flux $f_{DR}$. The terminal model has same mass, $M_{RR} = M_{DR}$, same angular momentum, $L_{RR} = L_{xx}$, and same magnetic flux, $f_{RR} = f_{DR}$. If so, then the fitting between the starting model and the terminal model of such a ``nonaxisymmetric DRS-to-RRS transition'' (NXDRT) can proceed through the magnetic perturbation parameter $h_{RR}$ of the terminal model. The equations 
\begin{equation} \label{eq:LRR}
L_{RR} = L_{xx} = I_{RR,11}(h_{RR},L_{xx}) \Omega_1 {\rm ,} 
\end{equation}
\begin{equation} \label{eq:TRR}
T_{RR}(h_{RR}) = \frac{1}{2} I_{RR,11}(h_{RR},L_{xx}) \Omega_1^2 {\rm ,} 
\end{equation}
give, for various values $h_{RR}$ of the magnetic perturbation parameter, the corresponding angular velocity component $\Omega_1$ along the $I_1$ principal axis (with moment of inertia $I_{RR,11}(h_{RR},L_{xx})$; details on the calculation of this quantity are given in \S \, 5) and the rotational kinetic energy $T_{RR}$. If, furthermore, NXDRT takes place under constant magnetic perturbation parameter, $h_{RR} = h_{DR}$ (which seems to be a plausible assumption), then the system of equations (\ref{eq:LRR})--(\ref{eq:TRR}) gives the proper values for $\Omega_1$, $T_{RR}$, and the period, $P_{RR} = 2 \pi / \Omega_1$, of the terminal model. Then an estimate of the timescale $t_{DRD}$ needed for the spin period to increase from the value $P_{xx}$ to the value $P_{RR}$ is
\begin{equation} \label{eq:TDRD}
t_{DRD} \simeq \frac{P_{RR}-P_{xx}}{\dot{P}_{now}} {\rm ;}
\end{equation}
and an estimate for the NXDRT power is
\begin{equation} \label{eq:TdotDRD}
\dot{T} \simeq - \frac{T_{xx} - T_{RR}}{t_{DRD}} \equiv 
               - \frac{D\!E_{DRD}}{t_{DRD}} {\rm \, .}
\end{equation} 
In view of TOV scenario, therefore, the spin-down problem is drastically simplified, since, as our numerical results reveal, the NXDRT power is $\sim 10^2$ times less than the classically estimated spin-down power of equation (\ref{eq:Tdot}).

Finally, the following remark is of interest. In order to keep computations within the two-dimensional framework of the so-called ``complex-plane iterative technique'' (CIT) (GP99, \S 3 and references therein), we simulate the terminal rigidly rotating models by a constant-mass sequence of magnetic models with zero rotation (i.e., with rotational perturbation parameter $\upsilon = 0$) along the $I_3$ principal axis. Apparently, all the members of this nonrotating magnetic constant-mass sequence are prolate along the $I_3$ principal axis. Thus, our computations retain acceptable reliability, in describing the phenomenon under consideration, as far as the component $\Omega_3$ of ${\bf \Omega}$ remains small with respect to the component $\Omega_1$, the turn-over angle $\chi$ remains near $\case{\pi}{2}$, and the angle $\gamma$ of the angular velocity direction with respect to the angular momentum direction remains small. For the perpendicular rotator we have $\Omega_3 = 0$, $\chi = 90 \arcdeg$, and $\gamma = 0 \arcdeg$; thus, the perpendicular rotator can be described reliably by the simulation adopted.

\section{Details on the overall numerical method}
The axisymmetric model, initially used in TOV scenario, is computed by (the two-dimensional) CIT. In a previous investigation (\cite[hereafter G91, \S 3]{gero91}), CIT has been developed and used for computing rotating visco-polytropic models; this term indicates that the differential rotation of such models results from the particular equations governing the viscosity of their material. CIT has been  extended for computing rotating white dwarfs (\cite[hereafter GP94]{gepa94}), and rotating white dwarfs with toroidal magnetic field (\cite[hereafter GP97]{gepa97}). Recently (GP99), CIT has been further extended for computing rotating white dwarfs distorted by both toroidal and poloidal magnetic fields. An almost complete set of quantities involved in the computations is given in GP99 (Table 1). As we have already clarified, in the present paper we shall use definitions and symbols identical to those in GP99. 

The algorithm of CIT is analyzed in G91 (\S \, 3; see also GP99, Fig. 1). The structure and the rotational behavior of the axisymmetric model is identical to that of GP94 (\S 2). The toroidal magnetic field, ${\bf H}_t = H_t {\bf u}_{\varphi}$, is defined as in GP97 (\S 2, eqs. [2.1]--[2.2]; see, also, GP99, eq. [2.7]); the poloidal magnetic field, ${\bf H}_p = H_{p \varpi} {\bf u}_{\varpi} + H_{pz} {\bf u}_z$, is defined as in GP99 (\S 2, eqs. [2.8]--[2.9]; $\varpi$, $\varphi$, and $z$ are the cylindrical coordinates). On the base of these definitions and, also, of the units and the corresponding dimensionless quantities defined in GP99 (Table 1), some straightforward algebra shows that
\begin{equation} \label{eq:Ht}
{\bf H}_t = H_t {\bf u}_{\varphi} = (k \varpi \varrho) {\bf u}_{\varphi} =
           [2 \pi \sqrt{G} \alpha B y_0^3] \sqrt{h} s 
           \left( \Phi^2 - \frac{1}{y_0^2} \right)^\frac{3}{2} 
           {\bf u}_{\varphi} {\rm ,} 
\end{equation} 
\begin{equation} \label{eq:Hp}
{\bf H}_p = H_{p \varpi} {\bf u}_{\varpi} + H_{pz} {\bf u}_z =
            \left( - \frac{k}{\beta} \varpi \frac{\partial \varrho}
            {\partial z} \right) {\bf u}_{\varpi} +
            \left( 2 \frac{k}{\beta} \varrho +
            \frac{k}{\beta} \varpi \frac{\partial \varrho}{\partial \varpi}
            \right) {\bf u}_z {\rm ,}
\end{equation}
\begin{equation} \label{eq:Hppi}
H_{p \varpi} = [2 \pi \sqrt{G} \alpha B y_0^3] \sqrt{\frac{h}{\beta_{*}^2}}
               \left( - 3 s \Phi \sqrt{\Phi^2 - \frac{1}{y_0^2}} 
               \frac{\partial \Phi}{\partial \zeta} \right) {\rm ,}
\end{equation}
\begin{equation} \label{eq:Hpzi}
H_{pz} = [2 \pi \sqrt{G} \alpha B y_0^3] \sqrt{\frac{h}{\beta_{*}^2}}
         \left( 2 \left( \Phi^2 - \frac{1}{y_0^2} \right)^\frac{3}{2} +
         3 s \Phi \sqrt{\Phi^2 - \frac{1}{y_0^2}} 
         \frac{\partial \Phi}{\partial s} \right) {\rm ,}
\end{equation}
where $s = \varpi / \alpha$ and $\zeta = z / \alpha$ are the dimensionless cylindrical coordinates ($\alpha$ is the well-known white-dwarf length unit). For configurations declining slightly from sphericity, we can use the approximative expressions
\begin{equation}
\frac{\partial \Phi}{\partial s} \simeq \sqrt{1- \nu^2} 
                                 \frac{\partial \Phi}{\partial \xi} {\rm ,}
\end{equation}
\begin{equation}
\frac{\partial \Phi}{\partial \zeta} \simeq \nu 
                                 \frac{\partial \Phi}{\partial \xi} {\rm ,}
\end{equation}
where $\xi = r / \alpha$ and $\nu = \cos \vartheta$ are the dimensionless modified spherical coordinates (taken together with the angle $\varphi$; on the other hand, $r$, $\vartheta$, and $\varphi$ are the usual spherical coordinates).

The average surface poloidal magnetic field
\begin{equation} \label{eq:Bs}
B_s \equiv \left< H_p(\xi_s,\nu) \right>_{\nu} = 
    \left\langle 
                \sqrt{H_{p \varpi}^2(\xi_s,\nu) + H_{pz}^2(\xi_s,\nu)} 
    \right\rangle_{\nu} {\rm \, ,}
\end{equation}
where the subscript $\nu$ denotes the average over this coordinate, is somewhat prescribed to the value $B_s^p$ by either the theory or the observations. So, there is need to define a ``surface zone'' with ``base'' at (dimensionless) radius $\xi_s$, say. For white-dwarf models, a common practice is to consider that the points below the boundary where electrons become degenerate --- i.e., the ``transition layer'' --- are located at distance $\xi_s$ from the center (see, e.g., \cite[hereafter ST83, \S 4.1]{shte83}). So, the transition layer coincides with the base of the surface zone. The temperature at this transition layer (which is in turn equal to the interior temperature), $\frak{T}_s$, is given in terms of the luminosity, $\frak{L}$, the mass, $M/M_{\sun}$, and the composition, $X$, $Y$, $Z$, appearing either explicitly or through the quantities
\begin{equation}
m = \frac{1}{2X+
           \frac{3}{4}Y+\frac{1}{2}Z} {\rm ,}
\end{equation}
\begin{equation}
m_e = \frac{1}{X+
             \frac{1}{2}Y+\frac{1}{2}Z} {\rm ,}
\end{equation}
where $m$ is the ``mean molecular weight'' and $m_e$ the ``mean molecular weight per free electron'', both measured in proton masses (see, e.g., ST83, \S 2.3). In particular, $\frak{T}_s$ is given by the relation (ST83, eq. [4.1.11])
\begin{equation} \label{eq:Ts}
\frak{T}_s = \left( \frac{Z(1+X) m_e^2 \frak{L}}
                        {5.7 \times 10^5 m \frac{M}{M_{\sun}}}
             \right)^{\frac{1}{3.5}} {\rm ,}
\end{equation}
with corresponding density at the transition layer (ST83, eq. [4.1.10])  
\begin{equation} \label{eq:rhos}
\varrho_s = 2.4 \times 10^{-8} m_e \frak{T}_s^{1.5} {\rm .}
\end{equation}
Thus, the radius $\xi_s$ results as root of the nonlinear algebraic equation
\begin{equation}
\varrho(\xi_s) - \varrho_s = 0 {\rm ,}
\end{equation}
or, equivalently,
\begin{equation}
By_0^3 \left( \Phi^2(\xi_s) - \frac{1}{y_0^2} \right)^{3/2} - 
                                                \varrho_s = 0 {\rm ,}
\end{equation}
which can be solved by a standard algorithm of numerical analysis. 

The fitting between the prescribed average surface poloidal magnetic field, $B_s^p$, and the ``proper poloidal magnetic parameter'', $\beta_{*}^p$, can proceed through the nonlinear algebraic equation 
\begin{equation}
B_s^p = \left\langle \sqrt{H_{p \varpi}^2(h;\beta_{*}^p;\xi_s,\nu) + 
                    H_{pz}^2(h;\beta_{*}^p;\xi_s,\nu)} \right\rangle_{\nu}                      {\rm ,}
\end{equation}
which, for prescribed $h$, can be solved by a standard algorithm of numerical analysis and give $\beta_{*}^p$ as root.     

Concerning the computation of constant-mass sequences, the white dwarf case is quite different to the polytropic case. In the polytropic case, one can construct a constant-mass sequence simply by keeping a particular fixed value for the fundamental model parameter $n$, i.e., for the polytropic index (see, e.g., \cite[\S 2]{shte90}; also, \cite[\S 3]{gall92}). On the other hand, a constant-mass sequence of white dwarfs consists of models with properly varying fundamental model parameter $y_0$ or, equivalently, $\frac{1}{y_0^2}$. This is due to the fact that the central density $\varrho_c$ depends explicitly on $y_0$ through the well-known relation 
\begin{equation}
\varrho_c = B y_0^3 \left( 1- \frac{1}{y_0^2} \right)^{3/2}.
\end{equation}
Thus, a rotating magnetic white dwarf may attain a specific mass only for a specific combination of the fundamental model parameter $y_0$ and the rotational perturbation parameter $\upsilon$.

It is worth emphasizing that white dwarfs with given $\frac{1}{y_0^2}$ may not be able to attain a specific mass at all, since there are lower and upper mass limits; namely, the lower mass is that of a nonrotating nonmagnetic model, and the upper mass (for given magnetic perturbation parameter $h$) is the mass of a critically rotating model.

In order to compute constant-mass sequences of white dwarfs, we use CIT in an iterative algorithm, called ``search for specific mass'' (SSM). This algorithm finds the value $\upsilon_{M,h}$ for which a model with given  $\frac{1}{y_0^2}$ and $h$ has a specific mass $M/M_{\sun}$. Alternatively, SSM can find the value $h_{M,\upsilon = 0}$ for which a nonrotating model (i.e., $\upsilon = 0$) with given $\frac{1}{y_0^2}$ has a specific mass $M/M_{\sun}$.  
Using SSM we can compute several constant-mass rotating sequences of white dwarfs. By calculating the angular velocity and the angular momentum of each member of a rotating sequence, we can plot the curve $\Omega(L)$ corresponding to a specific mass. However, in order to construct ``complete curves'', we need to know the extrema of the fundamental model parameter $\frac{1}{y_0^2}$, for which the sequence can still attain the specific mass. The maximum value $(\frac{1}{y_0^2})_{max}$ corresponds to the rightmost point of the curve $\Omega(L)$, that is, to an almost critically rotating model. A model with slightly larger $\frac{1}{y_0^2}$ will fail to attain the desired mass, since its mass is less than the desired one even in critical rotation. On the other hand, the minimum value $(\frac{1}{y_0^2})_{min}$ corresponds to the leftmost point of the curve, that is, to a model with rotational kinetic energy far from its critical value. A model with slightly smaller $\frac{1}{y_0^2}$ cannot attain the desired mass, since its lowest mass is greater than the desired one. 

To calculate the extrema of $\frac{1}{y_0^2}$ we have developed the so-called ``search for extrema of the basic model parameter'' (SEB). Starting from an intial guess for each extremum, this algorithm uses SSM iteratively to calculate the corresponding extrema. Alternatively, SEB can find the corresponding extrema of $\frac{1}{y_0^2}$ for a constant-mass nonrotating sequence, by properly adjusting the magnetic perturbation parameter $h$.   

To summarize, the computation of a rotating magnetic constant-mass sequence of white dwarfs consists of the following procedure. First, we specify certain values for the mass $M/M_{\sun}$, the magnetic perturbation parameter $h$, and the differential rotation reduction factor $F_r$. Alternatively, we take $\upsilon = 0$ --- nonrotating magnetic constant-mass sequence --- and we use in the fitting procedure a varying $h$. Second, we use SEB to calculate the corresponding $(\frac{1}{y_0^2})_{min}$ and $(\frac{1}{y_0^2})_{max}$. Third, we use SSM to compute several models within the already specified range of values. By these results we can study the behavior of the corresponding curve $\Omega(L)$. Alternatively, we can study the full range of values $(\frac{1}{y_0^2},h)$ for which a nonrotating magnetic sequence retains constant mass.

\section{Calculation of the moment of inertia of the perpendicular rotator along its rotation axis}
The quantity $I_{RR,11}(h_{RR},L_{xx})$, involved in equations (\ref{eq:LRR})--(\ref{eq:TRR}), denotes the moment of inertia of the perpendicular rotator along its rotation axis $I_1$. The problem of calculating this moment is three-dimensional, since the equilibrium structure of the perpendicular rotator is represented by an ellipsoidal figure. Thus, a systematic numerical treatment of this subject consists in developing an iterative  three-dimensional method, like that of \cite{hach86}, for computing equilibrium structures distorted by both rotation and magnetic field. Our plan (for the near future) is to modify and extend CIT in order to obtain the ``three-dimensional CIT'' (3DCIT), able to make the job required. At present, however, 3DCIT is not available. So, there is need to obtain a reliable estimate of $I_{RR,11}(h_{RR},L_{xx})$ by implementation of the existing ``two-dimensional CIT'' (2DCIT). 

We first clarify that 2DCIT can calculate (within a prescribed numerical accuracy) the moments $I_{11}(h_{RR})$ and $I_{33}(h_{RR})$ of a nonrotating magnetic constant-mass sequence, and the moment $I_{RR,33}(h_{RR},L_{xx})$ of an axisymmetric rigidly rotating magnetic constant-mass sequence. Indeed, in both cases the corresponding equilibrium structures are spheroidal figures; thus, they are involved in a two-dimensional problem easily handled by 2DCIT. 

Now, there is still a chance to obtain a reliable estimate of the so-called ``rigid rotation amplification ratio'', $A_r(I_1)$, resulting from the definition 
\begin{equation} \label{eq:Ar}
A_r(I_i) = \frac{I_{RR,ii}(h_{RR},L_{xx})}{I_{ii}(h_{RR})} 
\end{equation}
(with the substitutions $i = 1$, $ii = 11$), within the framework of 2DCIT. It is expected that $A_r(I_1) > 1$, since rotation does prolate further the configuration of the perpendicular rotator along the $I_3$ principal axis and, so, does increase the moment of inertia along the rotation axis $I_1$ from the value $I_{11}(h_{RR})$ (magnetic field only) to the value $I_{RR,11}(h_{RR},L_{xx})$ (magnetic field plus rigid rotation). To start with, let us consider our problem in view of a perturbation method, like that of \cite{ghad90}. In this view, the moments $I_{RR,11}(h_{RR},L_{xx})$ and $I_{RR,33}(h_{RR},L_{xx})$ are given by the expansions (see GH90, eqs. [3.6]--[3.8], [4.1], and other similar expansions therein)
\begin{equation}
I_{RR,ii}(h_{RR},L_{xx}) = I_{ii}(sph) + \Delta I_{ii}(mag) + \Delta I_{ii}(rot)                + \Delta I_{ii}(mag*rot) {\rm ,} 
\end{equation}
where the argument $sph$ shows the spherical (i.e., the nondistorted) contribution, the argument $mag$ gives distortion due to magnetic field, the argument $rot$ represents distortion due to rotation, and the argument $mag*rot$ shows mixed distorting action of both magnetic field and rotation. The significant point is that, when approximating the problem by a first-order perturbation method (see, e.g., \cite[\S\S \, 1--2]{ghad92}), then, apart from the spherical contribution, we take into account distortions due to magnetic field of first order in the magnetic perturbation parameter $h$, 
\begin{equation}
\Delta I_{ii}(mag) \simeq h M_{ii} {\rm ,}
\end{equation}
and distortions due to rotation of first order in the rotational perturbation parameter $\upsilon$, 
\begin{equation} \label{eq:DIiiR}
\Delta I_{ii}(rot) \simeq \upsilon R_{ii} {\rm ,}
\end{equation}
while distortions of order equal to or higher than $h^2$, $\upsilon^2$, $h \upsilon$ (mixed distorting terms), are omitted. It appears therefore that, within the framework of a first-order approximation, the distortions due to magnetic field and rotation become decoupled and the corresponding calculations 
\begin{equation} \label{eq:IRRii}
I_{RR,ii}(h_{RR},L_{xx}) \simeq I_{ii}(sph) + \Delta I_{ii}(mag) + \Delta I_{ii}(rot) 
\end{equation}
can be undertaken in separate steps concerning sph-terms, mag-terms, and rot-terms. Furthermore, combination of equation (\ref{eq:Ar}), defining $A_r(I_i)$, and of the approximating equation (\ref{eq:IRRii}) gives 
\begin{equation} \label{eq:ArIi}
A_r(I_i) \simeq \frac{I_{ii}(sph) + \Delta I_{ii}(mag) + \Delta I_{ii}(rot)}
                     {I_{ii}(sph) + \Delta I_{ii}(mag)} =
                 1 + \frac{\Delta I_{ii}(rot)}{I_{ii}(h_{RR})} {\rm ,}
\end{equation}
which says that either $A_r(I_i)$ is known and then $\Delta I_{ii}(rot)$ can be  calculated by 
\begin{equation} \label{eq:DIii}
\Delta I_{ii}(rot) \simeq \frac{A_r(I_i) -1}{I_{ii}(h_{RR})} {\rm ,}
\end{equation}
or $\Delta I_{ii}(rot)$ is known and then $A_r(I_i)$ can be calculated by equation (\ref{eq:ArIi}).

Now, there is an obvious ``invariance'' concerning rotational distortion with respect to the principal axes $I_1$, $I_2$, $I_3$, when departing from an initial spherical (i.e., nondistorted) configuration. In particular (and for same rotational perturbation parameter $\upsilon$), when taking $I_3$ as rotation axis, we calculate a distortion $\Delta I_{33}(rot)$ equal to the distortion $\Delta I_{22}(rot)$ calculated when taking $I_2$ as rotation axis, and, in turn, equal to the distortion $\Delta I_{11}(rot)$ calculated when taking $I_1$ as rotation axis. Because of this invariance, the distortion $\Delta I_{11}(rot)$ is equal to the distortion $\Delta I_{33}(rot)$ calculated by equation (\ref{eq:DIii}) (with the substitutions $i = 3$ and $ii = 33$). We remark that all the quantities involved in equations (\ref{eq:ArIi})--(\ref{eq:DIii}) can be calculated within the framework of 2DCIT; so, the moment of inertia $I_{RR,11}(h_{RR},L_{xx}) = A_r(I_1) I_{11}(h_{RR})$ can be eventually calculated by 2DCIT. 

The next significant point is that the angular velocity $\Omega_1$ of the perpendicular rotator,
\begin{equation} \label{eq:Wmega}
\Omega_1 = \frac{L_{xx}}{A_r(I_1)I_{11}(h_{RR})} {\rm ,}
\end{equation}
should give rotational perturbation parameter $\upsilon$,
\begin{equation} \label{eq:upsilon}
\upsilon = \frac{\Omega_1^2}{2 \pi G B y_0^3(h_{RR})} 
\end{equation}
(where the function $y_0 = y_0(h)$ is known from our nonrotating magnetic constant-mass sequence), equal to the value of $\upsilon$ used in equation (\ref{eq:DIii}) (with the substitutions $i = 3$ and $ii = 33$) for evaluating $\Delta I_{33}(rot)$, which is taken, in turn, from the known function $\upsilon = \upsilon(L)$ holding for our axisymmetric rigidly rotating magnetic constant-mass sequence. Accordingly, a fitting is required between the value of $\upsilon$ resulting from the approximation (\ref{eq:IRRii})--(\ref{eq:upsilon}) and the value $\upsilon = \upsilon(L)$ used for deriving this approximation. A simplified iterative algorithm undertaking this fitting can be constructed on the basis of the remark that the so-called ``rotational distortion coefficient'' $R_{ii}$ in equation (\ref{eq:DIiiR}) varies slow with the angular momentum, while $\upsilon$ varies fast. In other words, $R_{ii}$ can be assumed constant over a narrow range $[L(P_{now}),L_x]$ of angular momenta. In detail, this ``simplified algorithm'' (S) has as follows.

Step 1(S). Take $L = L_{RR} = L_{xx}$ for the angular momentum and $h = h_{RR} = h_{DR}$ for the magnetic perturbation parameter; then calculate the rotational perturbation parameter $\upsilon^{(1)}$ from the known function $\upsilon = \upsilon(L)$ holding for our axisymmetric rigidly rotating magnetic constant-mass sequence, $\upsilon^{(1)} = \upsilon(L_{xx})$.  

Step 2(S). Calculate the rigid rotation amplification ratio $A_r(I_3)$ from the relation (eq. [\ref{eq:Ar}] with $i = 3$ and $ii = 33$)  
\begin{equation}
A_r(I_3) = \frac{I_{RR,33}(h_{RR},L_{xx})}{I_{33}(h_{RR})} {\rm ,}
\end{equation}
the distortion $\Delta I_{33}(rot)$ from the relation (eq. [\ref{eq:DIii}] with $i = 3$ and $ii = 33$)
\begin{equation}
\Delta I_{33}(rot) \simeq \frac{A_r(I_3) -1}{I_{33}(h_{RR})} {\rm ,}
\end{equation}
and the rotational distortion coefficients $R_{11}$, $R_{33}$ from the relation (eq. [\ref{eq:DIiiR}] with $ii =33$) 
\begin{equation}
R_{11} = R_{33} = \frac{\Delta I_{33}(rot)}{\upsilon^{(1)}} {\rm .}
\end{equation}

Step 3(S). Calculate the distortion $\Delta I_{11}(rot)^{(1)}$ from the relation (eq. [\ref{eq:DIiiR}] with $ii = 11$) 
\begin{equation}
\Delta I_{11}(rot)^{(1)} = R_{11} \upsilon^{(1)} {\rm ,}
\end{equation}
the rigid rotation amplification ratio $A_r(I_1)^{(1)}$ from the relation (eq. [\ref{eq:Ar}] with $i = 1$ and $ii = 11$)
\begin{equation}
A_r(I_1)^{(1)} \simeq 1 + \frac{\Delta I_{11}(rot)^{(1)}}{I_{11}(h_{RR})} 
                               {\rm ,}
\end{equation}
and take $I_{RR,11}(h_{RR},L_{xx})^{(1)} = A_r(I_1)^{(1)} I_{11}(h_{RR})$; then calculate the angular velocity $\Omega_1^{(1)}$ from the relation (eq. [\ref{eq:Wmega}])
\begin{equation}
\Omega_1^{(1)} = \frac{L_{xx}}{I_{RR,11}(h_{RR},L_{xx})^{(1)}} {\rm ,}
\end{equation}
and the rotational perturbation parameter $\upsilon^{(2)}$ from the relation (eq. [\ref{eq:upsilon}]) 
\begin{equation}
\upsilon^{(2)} = \frac{(\Omega_1^{(1)})^2}{2 \pi G B y_0^3(h_{RR})} {\rm ,} 
\end{equation}
where the function $y = y_0(h)$ is known from our nonrotating magnetic constant-mass sequence. 

Step 4(S). Check if $|\upsilon^{(2)} - \upsilon^{(1)}| < tol$, where $tol$ is a proper tolerance. If yes, then $I_{RR,11}(h_{RR},L_{xx})^{(1)}$ is a satisfactory approximation of $I_{RR,11}(h_{RR},L_{xx})$. If no, then use the value $\upsilon = \upsilon^{(2)}$ for the next iteration, starting from Step 3(S). Remark that the iteration counter obtains now the value $(2)$ for Step 3(S), where, however, the new estimate for $\upsilon$ takes as iteration counter the value $(3)$, $\upsilon^{(3)}$. So, when arriving again at Step 4(S), the comparison is made between the values $\upsilon^{(3)}$ and $\upsilon^{(2)}$; etc.          

The ``complete algorithm'' (C), on the other hand, takes into account even the slow variation of the rotational distortion coefficient $R_{ii}$. This algorithm has as follows.

Step 1(C). Take $L^{(1)} = L_{RR} = L_{xx}$ for the angular momentum and $h = h_{RR} = h_{DR}$ for the magnetic perturbation parameter; then calculate the rotational perturbation parameter $\upsilon^{(1)}$ from the known function $\upsilon = \upsilon(L)$ holding for our axisymmetric rigidly rotating magnetic constant-mass sequence, $\upsilon^{(1)} = \upsilon(L^{(1)})$.  

Step 2(C). Calculate the rigid rotation amplification ratio $A_r(I_3)^{(1)}$ from the relation
\begin{equation}
A_r(I_3)^{(1)} = \frac{I_{RR,33}(h_{RR},L^{(1)})}{I_{33}(h_{RR})} {\rm ,}
\end{equation}
and the distortion $\Delta I_{33}(rot)^{(1)}$ from equation (\ref{eq:DIii}); then take $\Delta I_{11}(rot)^{(1)} = \Delta I_{33}(rot)^{(1)}$ and calculate the rigid rotation amplification ratio $A_r(I_1)^{(1)}$ from the relation
\begin{equation}
A_r(I_1)^{(1)} \simeq 1 + \frac{\Delta I_{11}(rot)^{(1)}}{I_{11}(h_{RR})} 
                               {\rm .}
\end{equation}

Step 3(C). Take $I_{RR,11}(h_{RR},L^{(1)}) = A_r(I_1)^{(1)} I_{11}(h_{RR})$; then calculate the angular velocity $\Omega_1^{(1)}$ from the relation
\begin{equation}
\Omega_1^{(1)} = \frac{L_{xx}}{I_{RR,11}(h_{RR},L^{(1)})} {\rm ,}
\end{equation}
and the rotational perturbation parameter $\upsilon^{(2)}$ from the relation 
\begin{equation}
\upsilon^{(2)} = \frac{(\Omega_1^{(1)})^2}{2 \pi G B y_0^3(h_{RR})} {\rm ,} 
\end{equation}
where the function $y_0 = y_0(h)$ is known from our nonrotating magnetic constant-mass sequence. 

Step 4(C). Check if $|\upsilon^{(2)} - \upsilon^{(1)}| < tol$, where $tol$ is a proper tolerance. If yes, then $I_{RR,11}(h_{RR},L^{(1)})$ is a satisfactory approximation of $I_{RR,11}(h_{RR},L_{xx})$. If no, then calculate the root $L^{(2)}$ of the algebraic equation $\upsilon(L) = \upsilon^{(2)}$, and use the value $L = L^{(2)}$ for the next iteration, starting from Step 2(C). Remark that the iteration counter obtains now the value $(2)$ for Steps 2(C) and 3(C), where, however, the new estimate for $\upsilon$ takes as iteration counter the value $(3)$, $\upsilon^{(3)}$. So, when arriving again at Step 4(C), the comparison is made between the values $\upsilon^{(3)}$ and $\upsilon^{(2)}$; etc. 

\section{Turn-over and differential rotation decrease in nonaxisymmetric transitions}
MHD studies of rotating stars have been undertaken by several investigators (see, e.g., \cite{chaa56}; \cite{chab56}; \cite{cken57}; \cite{wdav67}; \cite{park70}; \cite{cgab72}; MT72; \cite{mona73}; \cite{wrig73}; \cite{wrig74}; \cite{mmos77}; \cite[hereafter M77]{mosa77}; \cite{mosb77}; \cite{chan79}; \cite{dick79}; \cite{moss79}; \cite{moss81}; \cite{gwoo85}; \cite{ptay85}; \cite{trfi85}; \cite[hereafter TT86]{ttaa86}; \cite{ttab86}; \cite{mwei87}; \cite[hereafter MMT88]{mmta88}; \cite[hereafter TT89]{ttas89}; MMT90; \cite[hereafter CM92]{cmac92}; M92; \cite{cmac93}; \cite{mshi94}; \cite{mbbt95}; \cite{ttbm95}).

In case of a model undergoing ``axisymmetric DRS-to-RRS transition'' (XDRT), the effect of mutual interaction of mixed magnetic field and rotation has become an issue of long debate among investigators asserting different aspects. In particular, a central question under debate has been the following: is there, in general, an unobtrusive magnetic field that can enforce uniform rotation and laminar meridional streaming? First, the study by TT86 (\S \, 8) gives negative answer to this question, unless turbulent viscosity is also involved in the process. Second, however, MMT88 (\S \, 1, 4) argue that differential rotation --- unless it is such that all parts of any poloidal field line have same angular velocity (situation known as ``isorotation'') --- will produce a toroidal field from the poloidal field. So, if differential rotation is maintained, the toroidal field will build up steadily until either the field is able to diffuse through the matter or the Lorentz force becomes so strong that it reacts back on differential rotation. Eventually, unless the magnetic field is very weak, the influence of the Lorentz force will be so strong that only very slight differential rotation will be possible. Third, in turn, TT89 (\S \, 4) reconsider this matter and find that, although a fossil magnetic field may sometimes enforce almost uniform rotation at some places, it never acts uniformly in the whole stellar radiative envelope; instead, it always tends to generate large nonuniformities in both the angular velocity and the toroidal magnetic field --- unless a large eddy viscosity is assumed to participate in the process, which then counters the effects of the magnetic field. The apparently conflicting results of TT86 and TT89, on the one hand, and MMT88 (see also MMT90, \S \, 4), on the other, have been discussed in detail by CM92 (\S \, 4.4). The authors remark that it is the anchoring of all poloidal field lines in a rigidly rotating core that is ultimately responsible for the establishment of a state of near solid body rotation throughout the envelope. Therefore, the extend to which a weak poloidal field can maintain or enforce solid body rotation throughout the envelope depends critically on assumptions made regarding the behavior of the poloidal field at the boundaries (i.e., anchored field vs. unanchored field). However, they do also remark that even for fully core-anchored poloidal fields, quasi-steady differentially rotating configurations can sometimes materialize and survive for significant time intervals. 

In the present investigation, we focus our attention on NXDRTs only, mainly for three reasons; namely, when considering XDRTs (1) it is not clear how much strong should the magnetic field be so as to remove differential rotation, (2) it is not clear when and why do such transitions start, and (3) there should be strict axisymmetry in such models, since otherwise they could evolve to oblique rotators.

To begin with, an approximative estimate for the turn-over timescale, $t_{TOV}$, is given by (MT72, eq. [54]; see also some interesting remarks regarding differential rotation at the last two paragraphs of \S \, 3)
\begin{equation} \label{eq:ttov}
t_{TOV} \simeq 
  \frac{D\!E_{DRD}}{V}
  \left \langle \Omega_3 \right \rangle^{-2}
  \left \langle \frac{H^2}{8 \pi \varrho a^2} \right \rangle^{-2} 
  \left \langle \frac{1}{3} \, \frac{\varrho V_t}{\Lambda_t} \right \rangle_{sz}^{-1}
  \left \langle \frac{\Omega_3^2 r^2}{a^2} \right \rangle_{sz}^{-2}
  \left \langle \Lambda_t \right \rangle_{sz}^{-2} 
  {\rm \, ,} 
\end{equation}
where the subscript $sz$ denotes that the corresponding average is calculated over the surface zone with base at the transition layer, $\xi_s$, and top at the boundary, $\xi_1$, of the star; averages without subscripts are calculated over the whole star, i.e., they are global averages. This estimate results as the mean energy per unit volume available for dissipation, $D\!E_{DRD} / V$ ($D\!E_{DRD}$ is the excess rotational kinetic energy due to differential rotation, as in eq. [\ref{eq:TdotDRD}],  and $V$ is the volume of the configuration), divided by the mean energy dissipation  during the turn-over due to turbulent viscosity in the convective surface zone per unit volume and per unit time (i.e., the product of the last 5 terms in the right-hand side of eq. [\ref{eq:ttov}]; details on the derivation of such relation are given in MT72, eqs. [7], [49]--[50]). For the angular velocity, $\Omega_3(s(\xi,\nu))$, we adopt the improved steady state Clement's differential rotation model (angular velocity constant on cylinders; as in GP99, eq. [2.3])
\begin{equation}
\Omega_3(s) = \Omega_{3c} \omega(s) =
            \Omega_{3c} \sqrt{\sum_{i=1}^4{a_i e^{-F_r b_i s^2}}} {\rm \, ,}
\end{equation}
where $\Omega_{3c}$ is the central angular velocity along the $I_3$ principal axis, $\omega(s)$ is the differential rotation function, $a_i$ and $b_i$ are the so-called ``nonuniformity parameters'' (their values for some typical cases are given in \cite{ghad91}, Table 1), and $F_r$ is the so-called ``reduction factor'' (note that $\omega = 1$ when $F_r = 0$ [rigid rotation]; and $\omega = {\rm minimum}$ when $F_r = 1$ [situation known as ``physical differential rotation'']). For the magnetic field, ${\bf H}(\xi,\nu) = {\bf H}_t(\xi,\nu) + {\bf H}_p(\xi,\nu)$, we use equations (\ref{eq:Ht})--(\ref{eq:Hpzi}). The remaining quantities in equation (\ref{eq:ttov}) are calculated under the assumption of slightly declining from sphericity (assumption valid when calculating $t_{TOV}$ only). Accordingly, the density is approximated by $\varrho(\xi,\nu) \simeq \varrho(\xi)$, the pressure is approximated by $P(\xi,\nu) \simeq P(\xi)$, the local sound speed, $a$, is approximated by (see, e.g., \cite{mfon89}, \S \, 2.a)
\begin{equation}  
a(\xi,\nu) \simeq a(\xi) = \sqrt{\frac{P(\xi)}{\varrho(\xi)}} {\rm \, ,}
\end{equation}
the turbulent mixing length, $\Lambda_t$, results from the ``mixing-length theory'' (MLT) --- see, e.g., \cite[hereafter CM91,]{cmaz91} for interesting issues concerning MLT --- as $\Lambda_t = A_P h_P$ (CM91, \S \S \, 2.8, 3.2), where the fine-tuning parameter $A_P$ is taken to be of order unity, $A_P \simeq 1$, and $h_P$ is the pressure scaleheight, 
\begin{equation}
\Lambda_t(\xi,\nu) \simeq \Lambda_t(\xi) \simeq 
          h_P(\xi) = \frac{P(\xi)}{\varrho(\xi) g(\xi)} = 
          \frac{P(\xi)(\alpha \xi)^2}{G \varrho(\xi) M(\xi)} {\rm \, ;} 
\end{equation}
$M(\xi)$ is the mass inside a sphere of radius $\xi$, and the turbulent velocity, $V_t$, is approximated by (see, e.g., \cite[hereafter B90]{bric90}, eq. [5]; in combination with CM91, eq. [38])
\begin{equation}
V_t(\xi,\nu) \simeq V_t(\xi) \simeq 
             \sqrt{\frac{3 P(\xi)}{\varrho(\xi)}} {\rm \, .}
\end{equation}
It is worth remarking that, when adopting such approximations for $h_P$ and $V_t$, their average values over the surface zone for a  white dwarf with mass $M \simeq 0.9 M_{\sun}$, radius $R = \alpha \xi_1 \simeq 7 \times 10^8 \, \rm cm$,  and temperature $\frak{T}_s \simeq 4 \times 10^7 \, \rm K$ are, respectively,
\begin{equation}
\left\langle h_P \right\rangle_{sz} \simeq 
     \frac{\int_{\xi_s}^{\xi_1}{h_P(\xi)}d\xi}{(\xi_1 - \xi_s)} \simeq 
     3 \times 10^5 \, {\rm cm} \sim \frac{R}{10^3} {\rm \, ,}
\end{equation}
\begin{equation}
\left\langle V_t \right\rangle_{sz} \simeq 
     \frac{\int_{\xi_s}^{\xi_1}{V_t(\xi)}d\xi}{(\xi_1 - \xi_s)} \simeq 
     5 \times 10^7 \, {\rm cm \, s^{-1}}
\end{equation}
(compare with corresponding values given by \cite{fuji88}, \S \, 2.1; see also \cite{fuji93}, \S \, 4.1; and B90, \S \, 4).

Our numerical results show that, for model parameters chosen as above, the turn-over timescale is of order $t_{TOV} \sim 10^7 \, \rm y$. Thus, the magnetic poles move towards the equator (on trajectories coinciding with circumference quadrants, $2 \pi R / 4$) with an average speed, $V_{mp}$, given by 
\begin{equation}
V_{mp} = \frac{\frac{1}{2} \pi R}{t_{TOV}} \sim 10^{-6} \, {\rm cm \, s^{-1} .} 
\end{equation}

A question arising here is what kind of ``globally acting'' (i.e., acting over the whole star) ``turn-over viscosity'', $\mu_{TOV}$, should substitute the ``locally acting'' (i.e., acting over the star's surface zone) ``turbulent viscosity'', $\mu_t$, so as to induce energy dissipated over the time $t_{TOV}$, $D\!E_{TOV}$, equal to the difference $D\!E_{DRD} = T_{xx} - T_{RR}$ of the rotational kinetic energies of the starting and the terminal models. The reason for seeking a global viscosity, instead of the local turbulent viscosity, is that differential rotation, which causes this kind of viscosity, is a global property of the model.

A first idea is to substitute the local turbulent velocity field, $V_t$, by a global ``turn-over velocity field'', $V_{TOV} \sim R / t_{TOV}$, which is of same order of magnitude with $V_{mp}$, $V_{TOV} \sim V_{mp}$. Then the turbulent viscosity (see, e.g., B90, eq. [8])
\begin{equation}        
\mu_t(\xi) = \frac{1}{3} \varrho(\xi) V_t(\xi) A_P h_P(\xi) \simeq
             \frac{1}{3} \varrho(\xi) 
             \left\langle V_t \right\rangle_{sz}
             \left\langle h_P \right\rangle_{sz}
\end{equation}
(with $A_P = 1$ as above) can be substituted by the turn-over viscosity
\begin{equation}        
\mu_{TOV}(\xi) \simeq
   \frac{1}{3} \varrho(\xi) 
   \left\langle V_{TOV} \right\rangle A_{TOV}^h 
   \left\langle h_P \right\rangle \simeq
   \frac{1}{3} \varrho(\xi) \frac{A_{TOV}^h R^2}{10^2 \, t_{TOV}} {\rm ,}   
\end{equation}
(since for our model $\left \langle h_P \right \rangle \sim R / 10^2$) where $A_{TOV}^h$ is a fine-tuning parameter. Furthermore, the energy dissipated due to such turn-over viscous friction per unit time, $D_{TOV}$, is given by (see, e.g., \cite{gsid95}, eq. [7])
\begin{equation} \label{eq:DTOV}
D_{TOV} = 4 \pi \alpha^3 \Omega_{3c}^2 
   \int_0^1{
            \int_0^{
                    \xi_1}{\mu_{TOV}(\xi) \, \xi^4 (1 - \nu^2) 
            \left( \frac{d\omega}{ds} \right)_{s=s(\xi,\nu)}^2 d\xi
                   }
           } d\nu {\rm ,}
\end{equation}  
and assuming that $D_{TOV}$ remains constant over the time $t_{TOV}$ the energy, $D\!E_{TOV}$, dissipated due to such viscous friction over the time $t_{TOV}$ is equal to $t_{TOV} D_{TOV}$. Typical values for our model are $D\!E_{DRD} \sim 10^{47} \, \rm erg$ and, when taking $A_{TOV}^h = 1$, $D\!E_{TOV} \sim 10^{46} \, \rm erg$; thus, the required equality $D\!E_{TOV} = D\!E_{DRD}$ gives $A_{TOV}^h \sim 10$ as proper order of magnitude for this fine-tuning parameter. 

A second idea is to develop a phenomenological model involving further details on the time evolution of the turn-over process (for similar considerations see M77, mainly \S \S 3--4; see also \cite{moss90}, especially \S \S \, 5--6, and Figs. 9--10). In the so-called ``mixer-mixture model'' (MMM), we assume that the turning over configuration acts as a ``mixer'' on the ``mixture'' configuration, which remains almost axisymmetric with respect to the invariant angular momentum axis, i.e., the $z$ axis (the meaning of ``almost axisymmetric configuration'' is that the angle $\gamma$ between its angular velocity and its invariant angular momentum remains small: $\gamma < 1 \arcdeg$, say; apparently, $\gamma$ becomes zero after the termination of the turn-over; in the following, we shall assume that such a deviation of the mixture configuration from axisymmetry is negligible and we shall treat it as being axisymmetric with respect to the $z$ axis). The mixer consumes its excess rotational kinetic energy due to differential rotation in the process of homogenizing the rotation of the mixture. In particular, when viewed in the mixture's system of reference, the mixer generates a velocity field, which carries mass elements across cylinders of different angular velocities and momenta. This is due to the fact that the rotational circular orbit of a mass element of the mixer, as seen in the mixer's system of reference, crosses several cylinders with different angular velocities, as seen in the mixture's system of reference. Thus, there is an interchange of angular momentum, which leads to uniform rotation of the mixture. 

For (1) the turn-over angle $\chi$ (with $\nu_{\chi} = cos(\chi)$), (2) the central angular velocity $\Omega_{3c}$ of the mixer, and (3) the differential rotation reduction factor $F_r$ of the mixer, we adopt the time dependent scheme 
\begin{equation} \label{eq:tdep}
Q(Q_{ini},t) = \left(
             \frac{1+e^{(\kappa -1)\delta}}
                  {1+e^{(\kappa - \nu_L(t))\delta}} 
       \right) Q_0(Q_{ini}) -
       \left(
             \frac{1+e^{(\kappa -1)\delta}}
                  {1+e^{\kappa \delta}} 
       \right) Q_0(Q_{ini}) {\rm ,}
\end{equation}     
where the so-called ``threshold parameter'', $\kappa$, is taken equal to $\kappa = cos(45 \arcdeg) \simeq 0.707$, the so-called ``delta parameter'', $\delta$, is chosen equal to $\delta \simeq 5$, and the auxiliary function $\nu_L(t)$ varies linearly with the time in the interval $[0,t_{TOV}]$,
\begin{equation} \label{eq:nuLt}
\nu_L(t) = \left \{ 
           \begin{array}{ll}
           1 - \frac{t}{t_{TOV}} & {\rm if} \ t \leq t_{TOV} \\
           0                     & {\rm if} \ t > t_{TOV}
           \end{array}
           \right. {\rm ;}
\end{equation}
When $\delta$ is relatively large ($\delta = 50$, say), the second term in the right-hand side of equation (\ref{eq:tdep}) becomes negligible with respect to its counterpart, and $Q_0(Q_{ini})$ can be simply set equal to the initial value(s) $Q_{ini}$, $\nu_{\chi[ini]} \lesssim 1$, $\Omega_{3c[ini]} = \Omega(L_{xx})$, and $F_{r[ini]} = 1.00, \; {\rm or} \; 0.50$. However, as our numerical results show, the present study requires $\delta \simeq 5$; since then the second term in the right-hand side of equation (\ref{eq:tdep}) cannot be neglected, $Q_0(Q_{ini})$ must be normalized to the value   
\begin{equation}
Q_0(Q_{ini}) = \frac{Q_{ini}}{1 - \left(
                         \frac{1+e^{(\kappa -1)\delta}}
                              {1+e^{\kappa \delta}}
                         \right)} {\rm \, ,}
\end{equation}
so that $Q(Q_{ini},0) = Q_{ini}$.
 
MT72 (\S \S \, 2--3) use a linear perturbation analysis to describe the field of motions, $\xi$ (MT72, eq. [44]), due to turn-over. In the framework of MMM, instead, we use a simplifying description according to which the turn-over speed at a particular point is proportional to the relative difference of the linear speeds due to differential rotation, $v_{mxr}(\xi, \nu_{mxr},t)$ and $v_{mxt}(\xi,\nu_{mxt},t)$, in the mixer and in the mixture (where the rotation of the latter results as superposition of a progressively established rigid rotation about the $z$ axis with angular velocity $\Omega_{z}$ [which is not involved in the calculation of the turn-over speed] and of a differential rotation about the $z$ axis identical to the differential rotation of the mixer about the $I_3$ principal axis; for simplicity, we shall use the same symbols for the central angular velocity $\Omega_{3c}$ and the angular velocity field $\Omega_3$ for the differential rotations of both the mixer and the mixture, keeping in mind that the only difference concerns their rotation axes),      
\begin{eqnarray} \label{eq:vtov}
V_{TOV} & = & V_{mp} 
              \left( 
              \frac{\left| v_{mxt}(\xi,\nu_{mxt},t) - 
                           v_{mxr}(\xi,\nu_{mxr},t) \right|}
                   {v_{mxt}(\xi,\nu_{mxt},t)}
              \right)  = \nonumber \\
        &   & V_{mp}
              \left(
              \frac{\left|
                    \omega(s(\xi,\nu_{mxt}),F_r(t)) s(\xi,\nu_{mxt}) -
                    \omega(s(\xi,\nu_{mxr}),F_r(t)) s(\xi,\nu_{mxr})
                    \right|}
                   {\omega(s(\xi,\nu_{mxt}),F_r(t)) s(\xi,\nu_{mxt})}
              \right) {\rm \, ,} \nonumber \\
        &   &   
\end{eqnarray}
where $V_{mp}$ is the expected order of magnitude for $V_{TOV}$. Regarding the relation between the coordinates $\nu_{mxt}$ and $\nu_{mxr}$ of a particular point, some elementary algebra (where the axisymmetry of the mixture configuration with respect to the $z$ axis is taken into account) shows that 
\begin{equation}
\vartheta_{mxt}(\vartheta_{mxr},\chi) = \left \{   
                  \begin{array}{ll}
   \chi + \vartheta_{mxr} & {\rm if} \ \chi + \vartheta_{mxr} \leq 90 \arcdeg \\
   180 \arcdeg - (\chi + \vartheta_{mxr}) & 
                          {\rm if} \ \chi + \vartheta_{mxr} > 90 \arcdeg 
                  \end{array} 
                  \right. {\rm ,}
\end{equation} 
and, accordingly,
\begin{equation}
\nu_{mxt}(\nu_{mxr},\nu_{\chi}) = \left \{ 
            \begin{array}{ll}
   + \nu_{\chi} \nu_{mxr} - \sqrt{1 - \nu_{\chi}^2} \sqrt{1 - \nu_{mxr}^2} &
                            {\rm if} \ \chi + \vartheta_{mxr} \leq 90 \arcdeg \\   
   - \nu_{\chi} \nu_{mxr} + \sqrt{1 - \nu_{\chi}^2} \sqrt{1 - \nu_{mxr}^2} &
                            {\rm if} \ \chi + \vartheta_{mxr} > 90 \arcdeg 
            \end{array}
            \right.
\end{equation}
Assuming for simplicity that the coordinate $\nu$ (without subscript) refers to the mixer configuration, we can write for the turn-over viscosity 
\begin{equation}        
\mu_{TOV}(\xi,\nu,t) \simeq \frac{1}{3} \varrho(\xi) 
              A_{TOV}^v V_{TOV}(\xi,\nu,t) A_{TOV}^h h_P(\xi) {\rm \, .}   
\end{equation}
Then the energy dissipated over the time $t_{TOV}$, $D\!E_{TOV}$, is given by
\begin{eqnarray} \label{eq:tdottov}
D\!E_{TOV}&=& \int_0^{t_{TOV}}{ D_{TOV}(t) } \, dt = 4 \pi \alpha^3 \times 
              \nonumber \\ 
          & & \int_0^{t_{TOV}}{ \Omega_{3c}^2(t) \left( 
                    \int_0^1{
                    \int_0^{\xi_1}{ \mu_{TOV}(\xi,\nu,t) \, \xi^4 (1 - \nu^2) 
                \left( \frac{\partial \omega(s,F_r(t))}{\partial s} 
                \right)_{s=s(\xi,\nu)}^2 d\xi }
                              } d\nu \right)} dt {\rm \, ;} 
              \nonumber \\
          & &
\end{eqnarray}  
remark that $D_{TOV}(t)$ is now a function of time, while in the foregoing interpretation it was taken approximately constant over $t_{TOV}$. Setting $A_{TOV}^h \sim 10$ as above, and $A_{TOV}^v =1$, we find $D\!E_{DRD} \sim 10^{47} \, \rm erg$ and $D\!E_{TOV} \sim 5 \times 10^{45} \, \rm erg$; thus, the required equality $D\!E_{TOV} = D\!E_{DRD}$ yields $A_{TOV}^v \sim 2 \times 10$ as proper order of magnitude for this fine-tuning parameter.

Eventually, we can adopt the above estimated values for the fine-tuning parameters, $A_P = 1$, $A_{TOV}^h = 10$, $A_{TOV}^v = 2 \times 10$, and try to find ``optimal values'' for the model parameters $L_{xx}$ (i.e., the invariant value of angular momentum under which turn-over takes place), $B_s$ (i.e., the average surface poloidal magnetic field), and $\delta$ (i.e., the delta parameter of the time dependent scheme [\ref{eq:tdep}]). These optimal values result as the roots of the nonlinear system of equations 
\begin{equation} \label{eq:optvals}
\left \{
\begin{array}{cccc}
t_{DRD} &=& t_{TOV} &  
  \textrm{i.e., equality imposed on relations (\ref{eq:TDRD}) and (\ref{eq:ttov}),} 
\\
D\!E_{DRD} &=& D\!E_{TOV} & 
  \textrm{i.e., equality imposed on relations 
                                        (\ref{eq:TdotDRD}) and (\ref{eq:tdottov}),}
\\
D\!E_{DRD} / t_{DRD} &=& D_{TOV}(t_{now}) & 
  \textrm{,}    
\end{array} 
\right.
\end{equation}      
which can be solved by a standard algorithm of numerical analysis. The last equation of this system has the physical meaning that we observe at the present time $t_{now}$ (such that $P(t_{now}) = P_{now}$) an energy dissipation time rate, $D_{TOV}(t_{now})$, more or less equal to its average value given by equation (\ref{eq:TdotDRD}); in other words, we assume that the turn-over observed now is smooth and described by its average values.

Finally, it is worth mentioning that the energy dissipation time rate due to the electron and ion viscosities of degenerate matter (see, e.g., \cite{duri73}, \S \, 3, eqs. [4], [6], respectively), $D_{EIV}$ (calculated by eq. [\ref{eq:DTOV}] with $\mu_{TOV}$ substituted by the electron--ion viscosity $\mu_{e+i}$), is $\sim 2$ orders of magnitude less than $D_{TOV}$; so, the former is totally screened by the latter and becomes insignificant in the framework of TOV scenario. 

\section{Is the magnetic field strong enough to remove differential rotation?}
A condition stating when the magnetic field dominates over the differential rotation, and eventually removes it, has been set up for a perpendicular rotator by MMT90 (\S \S \, 3--4; see also M92, \S 6). This condition involves the magnetic Reynolds number $p_1$ (MMT90, eq. [3.8a]),
\begin{equation}
p_1 = \frac{\nu_m}{\Omega R^2} {\rm \, ,}
\end{equation}
and a measure $p_2$ of the ratio $V_{As}^2 / V_{ROT}^2$ (where $V_{As} = B_s / \sqrt{4 \pi \varrho_s}$ is the well-known Alfv\'en speed at the base of the surface zone [see, e.g., CM92, eq. [13a]] and $V_{ROT}$ is the linear velocity due to rotation),
\begin{equation}
p_2 = \frac{B_s^2}{4 \pi \varrho_s \Omega^2 R^2} {\rm \, .}
\end{equation}
In these definitions, $\nu_m$ is the coefficient of magnetic diffusivity (see, e.g., \cite[hereafter T78]{tass78}, Chapt. 3, eq. [125]; see also ST83, eq. [7.1.8]),
\begin{equation}
\nu_m = \frac{c^2}{4 \pi \sigma} {\rm \, ,}
\end{equation}
where $\sigma$ is the coefficient of electrical conductivity (see, e.g., T78, Chapt. 3, eq. [120]; see also ST83, eq. [7.1.6]); $\Omega$ is taken here as the average value $\Omega = (\Omega_c + \Omega_e) / 2 = \Omega_c (1 + \omega_e) / 2$, where $\omega_e = \omega(s=\xi_e)$ is the differential rotation at the equatorial surface; $B_s$ is the average surface poloidal magnetic field (as in eq. [\ref{eq:Bs}]); $\varrho_s$ is the density at the transition layer (as in eq. [\ref{eq:rhos}]); and $R$ is taken here as the average value $R = (R_e + R_p) / 2 = \alpha (\xi_e + \xi_p) /2$, where $\xi_e$, $\xi_p$ are the dimensionless equatorial and polar radii, respectively. 

Using an argument set up earlier by \cite{radl86}, MMT90 show that, in order for the magnetic field to dominate over the differential rotation in case of a perpendicular rotator, $p_2$ must satisfy the condition (MMT90, eq. [4.1])
\begin{equation} \label{eq:p2gg}
p_2 \gg p_{2c} \simeq \frac{p_1^{2/3}}{4 \pi} 
    \left(
    \frac{\Delta \! \Omega}{\Omega}
    \right)^{4/3} {\rm \, ,}
\end{equation}
where $p_{2c}$ denotes a critical value for $p_2$, equal to the right-hand side of equation (\ref{eq:p2gg}), and $\Delta \! \Omega \simeq \Omega_c - \Omega_e = \Omega_c (1 - \omega_e)$. After some straightforward algebra, we find that this condition can be further written as 
\begin{equation} \label{eq:sigmagg}
\sigma \gg \sigma_c \simeq \frac{c^2 \left( \frac{1}{\omega_e} - 1 \right)^2}
  {(4 \pi)^{2.5} \left( \frac{1}{2} \Omega_c \left( 1 + \omega_e \right) \right)
   R^2 p_2^{1.5}} {\rm \, ,}
\end{equation}  
where $\sigma_c$ denotes a critical value for the electrical conductivity, equal to the right-hand side of equation (\ref{eq:sigmagg}); so, numerical study of the latter condition requires calculation of the electrical conductivity $\sigma$ of the model under consideration.

Electrical conductivity of compact stars has been studied by several investigators (see, e.g., \cite{fito76}; \cite[hereafter I\&83]{imii83}; \cite{ikoh93}; \cite[hereafter IHK93]{ihko93}; \cite{pote96}; \cite{pyak96}; \cite{pbhy99}; \cite[hereafter P99]{pote99}).When ignoring the influence of magnetic field on electrical conductivity, we can proceed to the calculation of $\sigma$ by constructing a software package based, e.g., on I\&83 (eqs. [2], [4]--[6], [8]--[9], and data from Table 1) and on IHK93 (eqs. [5], [7], [9], [28], [35], [39], [41], [43]--[44], [47], and data from Tables 1--3). For model parameters chosen as in \S \, 6, typical values concerning one-component plasmas (e.g., He plasma with atomic number $Z=2$ and mass number $A=4$, C plasma with $Z=6$ and $A=12$, O plasma with $Z=8$ and $A=16$, etc.) are $\sigma \sim 10^{19} \div 10^{20} \rm \, s^{-1}$, calculated at the base of the surface zone, and $\sigma_c \sim 10^{15} \rm \, s^{-1}$. So,  the condition (\ref{eq:sigmagg}) is easily satisfied by our model, since $\sigma / \sigma_c \sim 10^4 \div 10^5$. 

The main difference between the above study for a perpendicular rotator and the study for a turning over model (made in the framework of MMM) is that in the first case differential rotation is removed exclusively when the model is already perpendicular rotator; while in the second case differential rotation is progressively removed as the turn-over angle increases, and when the model becomes perpendicular rotator its rotation is already rigid. So, there is need to examine in detail the reasons which lead to angular momentum mixing in both models. 

In particular, angular momentum mixing takes place mainly due to the action of the Alfv\'en speed along the $\varpi_{mxt}$ direction (remember that $\Omega_3$ changes most rapidly along exactly this direction), $V_{A\varpi[mxt]}$ (generalization as in \cite{chan81}, Chapt. IV, eq. [60]), which carries angular momentum via hydromagnetic transverse Alfv\'en waves propagating along the poloidal magnetic component on this direction. In the perpendicular rotator, the poloidal magnetic component on the $\varpi_{mxt}$ direction, inducing $V_{A\varpi[mxt]}$, is the $z_{mxr}$-component $H_{pz[mxr]}$. Remark that the magnetic field remains axisymmetric in the mixer's system of reference; thus, equations (\ref{eq:Ht})--(\ref{eq:Hpzi}) for the magnetic field are expressed in terms of the mixer's coordinates, $\xi_{mxr} = \xi_{mxt} ( = \xi {\rm , \ for \ simplicity})$, and
\begin{equation}
\nu_{mxr}(\nu_{mxt},\nu_{\chi}) = \nu_{mxt} \, \nu_{\chi} + 
          \sqrt{1 - \nu_{mxt}^2} \, \sqrt{1 - \nu_{\chi}^2} {\rm \, .}
\end{equation}
In the starting model, on the other hand, the poloidal magnetic component inducing $V_{A\varpi[mxt]}$ is the $\varpi_{mxr}$-component $H_{p\varpi[mxr]}$. Generally, in any intermediate configuration (oblique rotator) the corresponding poloidal magnetic component, $H_{p\varpi[mxt]}$, is given by
\begin{eqnarray} \label{eq:Hpdirpi}
H_{p\varpi[mxt]}(\xi,\nu_{mxt},\nu_{\chi}) & = &
       {\bf H}_p \cdot {\bf u}_{\varpi[mxt]} = \nonumber \\
  & & H_{p\varpi[mxr]}(\xi,\nu_{mxr}(\nu_{mxt},\nu_{\chi})) \, \nu_{\chi} +  
      H_{pz[mxr]}(\xi,\nu_{mxr}(\nu_{mxt},\nu_{\chi})) \sqrt{1 - \nu_{\chi}^2} 
      {\rm \, ;} \nonumber \\
  & & 
\end{eqnarray}         
so, when $\nu_{\chi} \simeq 0$ (perpendicular rotator) we verify the first case, and when $\nu_{\chi} \lesssim 1$ (starting model with $\chi \gtrsim 0 \arcdeg$) we verify the second case.

Now, the difficulty with the starting model has to do with the fact that, for model parameters chosen as in \S \, 6, 
\begin{equation}
\frac{\left \langle H_{pz[mxr]} \right \rangle}
     {\left \langle H_{p\varpi[mxr]} \right \rangle} \sim 10 {\rm \, ;}
\end{equation} 
in other words, angular momentum mixing in the starting model should be carried out by the ``weak'' poloidal magnetic component, $H_{p\varpi[mxr]}$. Then, after turn-over has started (assuming that $H_{p\varpi[mxr]}$ can indeed drive angular momentum mixing), the magnetic field succeeds in progressively removing differential rotation for any intermediate configuration, since (1) angular momentum mixing is then carried out by the component $H_{p\varpi[mxt]} \, (> H_{p\varpi[mxr]})$, and (2) $\Omega_{3c}(t)$ and $F_r(t)$ decrease as $t$ increases (thus, as seen by eq. [\ref{eq:sigmagg}], $\sigma_c$ decreases as $t$ increases). 
   
On the other hand, magnetic fields in compact stars complicate electron transport making it particularly anisotropic (see, e.g., P99, \S \, 1). As a result, electrical conductivity becomes a second-rank tensor (P99, eqs. [16]--[18]) with prevailing components the longitudinal electrical conductivity, $\sigma_\parallel$, acting along the resultant field lines, and the transverse electrical conductivity, $\sigma_\perp$, acting across the resultant field lines. To take into account such effects, we introduce a two-fold generalization. First, the electrical conductivities of interest are now written as
\begin{equation}
\sigma_\parallel = \sigma_\parallel({\bf H}) = 
                   \sigma_\parallel({\bf H}_t + {\bf H}_p) \simeq
                   \sigma_\parallel({\bf H }_t) {\rm \, ,}
\end{equation}
\begin{equation}
\sigma_\perp = \sigma_\perp({\bf H}) =
               \sigma_\perp({\bf H}_t + {\bf H}_p) \simeq
               \sigma_\perp({\bf H}_t) 
\end{equation}
(magnetic fields as in eqs. [\ref{eq:Ht}]--[\ref{eq:Hpzi}]), since the ratio $H_t / H_p$ is large over the whole star, except for two confined regions near the center and near the boundary; typical values for our model are $H_t / H_p \sim 10^3 \div 10^4$. Now, in the starting model, angular momentum mixing is mainly carried out by hydromagnetic Alfv\'en waves propagating along the field lines of the poloidal magnetic field, that is, across the field lines of the resultant magnetic field, ${\bf H}$, which, in turn, does almost coincide with the toroidal field, ${\bf H}_t$. Remark that, for the starting model, the corresponding Alfv\'en waves propagating along the resultant field lines do not contribute to the mixing process since, by this way, they are trapped to travel on cylinders of constant angular velocity. Accordingly, angular momentum mixing is mainly monitored by the transverse electrical conductivity of the toroidal magnetic field, $\sigma_\perp(H_t)$. Thus, in summary, in the starting model angular momentum mixing is powered by $H_{p\varpi[mxt]}$ (as discussed above) and its efficiency is monitored by $\sigma_\perp(H_t)$ (as discussed here).

Second, the quantities $p_2$ and $\sigma_c$ become now functions of position and time (in the rest of this section, we shall use the mixture's system of reference; and, for simplicity, we shall denote the coordinate $\nu_{mxt}$ by $\nu$ alone), 
\begin{equation}
p_2(\xi,\nu,t) = \frac{H_{p\varpi[mxt]}^2(\xi,\nu,\nu_{\chi}(t))}
   {4 \pi \varrho(\xi,\nu) \, 
    \Omega_{3c}^2(t) \, \omega^2(s(\xi,\nu),F_r(t)) \, 
                                \alpha^2 s^2(\xi,\nu)} {\rm \, ,}
\end{equation}
\begin{equation}
\sigma_c(\xi,\nu,t) = 
   \frac{c^2 \left( \frac{1}{\omega(s(\xi,\nu),F_r(t))} -1 \right)^2}
   {(4 \pi)^{2.5} \, \Omega_{3c}(t) \, 
             \omega(s(\xi,\nu),F_r(t)) \, \alpha^2 s^2(\xi,\nu) \, 
                                     p_2^{1.5}(\xi,\nu,t)} {\rm \, ,}
\end{equation}
where, in addition, we have substituted the poloidal field $H_p$ by its component $H_{p\varpi[mxt]}$ along the $\varpi_{mxt}$ direction, since we are interested for the corresponding Alfv\'en speed $V_{A\varpi[mxt]}$, which, as discussed above, has maximum efficiency in interchanging angular momentum.

On the basis of such a two-fold generalization, the condition (\ref{eq:sigmagg}) is now written as 
\begin{equation} \label{eq:sigmaggnew}
\left \langle 
      \frac{\sigma_\perp(H_t(\xi,\nu_{mxr}(\nu,\nu_{\chi}(t))))}
           {\sigma_c(\xi,\nu,t)} 
      \right \rangle \gg 1 {\rm \, .}
\end{equation}
To compute the transverse electrical conductivity $\sigma_\perp(H_t)$, we use the Fortran Code {\tt CONDUCT}, downloaded from the site {\tt http://www.ioffe.rssi.ru/astro/conduct}. This computer code has been written by Potekhin and, actually, is the code that implements the formulae set up in P99 (see note in P99, last paragraph of \S \, 6). Typical values for the starting model (with model parameters chosen as in \S \, 6), when setting $\chi(0) = \chi_{ini} = \pi / 128 \simeq 1.41 \arcdeg$ (which gives in turn $\nu_{\chi}(0) = \nu_{\chi[ini]} = 0.9997$) and when taking as typical case a carbon plasma, are $\left \langle \sigma_\perp(H_t) / \sigma_c \right \rangle \sim 10^3$.

Such values, although apparently satisfying the condition (\ref{eq:sigmaggnew}), reveal a critical issue concerning the starting model. In particular, it seems that the toroidal magnetic field must be strong enough so as to induce an adequate dynamical asymmetry for the turn-over to take place; while, at the same time, it must be weak enough so as to permit interchanging of angular momentum via hydromagnetic Alfv\'en waves propagating across its field lines. The difficulty is that, on the one hand, dynamical asymmetry increases with the toroidal magnetic field (i.e., with the magnetic perturbation parameter $h$) and, on the other hand, efficiency in interchanging angular momentum decreases as $h$ increases (that is, $\sigma_\perp(H_t)$ decreases as h increases). In view of this remark and to the extent that our problem concerns mainly the behavior of the starting model, the right title for the present section seems to be: ``is the toroidal magnetic magnetic field strong enough to induce dynamical asymmetry and weak enough to permit angular momentum mixing?''.

\section{Results and discussion}
In our computations, we first choose $M = 0.89 M_{\sun}$ and $F_r =1.00$ (physical differential rotation). Taking then as typical case a carbon plasma with luminosity $\frak{L} \simeq 10^{32} \rm \, erg \, s^{-1}$ (which gives in turn interior temperature $\frak{T}_s \simeq 3.97 \times 10^7 \, \rm K$, calculated by eq. [\ref{eq:Ts}]), and assuming for the starting model $\chi(0) = \chi_{ini} = \pi / 128 \simeq 1.41 \arcdeg$ (which gives in turn $\nu_{\chi}(0) = \nu_{\chi[ini]} = 0.9997$), we compute the quantities $(I_{11}/I_{33} -1) \times 10$, which is a measure of the dynamical asymmetry, and $\left \langle \sigma_\perp(H_t) / \sigma_c \right \rangle \times 10^{-3}$, which is a measure of the degree that toroidal magnetic field permits angular momentum mixing, in terms of $h$. The computations are performed with angular momentum $L_{xx} \simeq (L(P_{now}) + L_x)/2$ (symbols as in \S \S \, 3--4). 

Figure 1 shows these two quantities as functions of $h$. We remark that an ``optimal value'' for the magnetic perturbation parameter is $h \simeq 0.08$ in the sense that at this value the ratio $\left \langle \sigma_\perp(H_t) / \sigma_c \right \rangle \times 10^{-3} \simeq 1.1$ is still greater than unity, while at the same time the quantity $(I_{11}/I_{33} -1) \times 10 \simeq 0.3$ verifies that dynamical asymmetry has been already set up. We therefore take $h_{DR} = h_{RR} = h = 0.08$ for all the subsequent computations.

For the constant-mass sequence $[M = 0.89 M_{\sun}, h=0.08, F_r =1.00]$, Figure 2 shows the graph of the function $P = P(L)$. The point $a(L=0.979 \times 10^{49} {\rm \, erg \, s}, P=52.091 {\rm \, s})$ represents an almost spherical model due to counterbalancing of the effects of rotation and poloidal field (both tending to derive oblate configurations) with the effects of toroidal field (tending in turn to form prolate configurations); the effective ellipticity of this model is $|e_{eff}| = |(\xi_{av} - \xi_p)| / \xi_{av} < 10^{-3}$. The point $b(L=1.615 \times 10^{49} {\rm \, erg \, s}, P=33.080 {\rm \, s})$ is a model with period equal to $P_{now}$; so, $L(P_{now}) = L_b = 1.615 \times 10^{49} {\rm \, erg \, s}$. The point $c(L=2.519 \times 10^{49} {\rm \, erg \, s},  P=22.150 {\rm \, s})$ is the model for which $I_{11} = I_{33}$ (dynamical asymmetry holds left to this point); so, $L_{x} = L_c = 2.519 \times 10^{49} \rm \, erg \, s$. The point $d(L=7.904 \times 10^{49} {\rm \, erg \, s}, P=12.634 {\rm \, s})$ is the model of minimum period, $P_{min} = 12.634 \rm \, s$; so, $L(P_{min}) = L_d = 7.904 \times 10^{49} \rm \, erg \, s$. This point defines two branches on the graph; in particular, on the branch left to $d$, secular evolution of the star induces a spin-down; while on the branch right to $d$ (that is, evolution from $e$ to $d$), secular evolution induces a spin-up. As in a recent paper (\cite[hereafter GP00, \S \, 4]{gpap00}), we assume that secular evolution of the star is initiated at a point left to $d$ and almost coinciding with it; eventually, this assumption means that the star does not undergo any secular spin-up during its evolution. The secular spin-down time rate, $\dot{P}_{sec}$, is involved in the relation (GP00, eq. [3])
\begin{equation} \label{eq:BsPs}
B_s \simeq \left( \frac{3c^3}{8\pi^2} I R_{av}^{-6} 
           \dot{P}_{sec} P_{sec} \right)^{\frac{1}{2}} {\rm \, ,} 
\end{equation}
which yields as typical values for our model $\dot{P}_{sec} \sim 10^{-3} \div 10^{-2} \dot{P}_{now}$ with angular momentum lying in the interval $[L_b,L_{x}]$; in other words, the secular spin-down time rate is totally screened by the spin-down time rate observed now, as suggested in the first paragraph of \S \, 3. Finally, the rightmost point $e(L=16.553 \times 10^{49} {\rm \, erg \, s}, P=16.582 {\rm \, s})$ represents a critically rotating model with ratio of kinetic to gravitational energy $\tau = T / |W| \simeq 0.14$ and effective ellipticity $e_{eff} \simeq 0.39$.

Numerical solution of the system (\ref{eq:optvals}) gives as optimal values for our model $L_{xx} = 2.154 \times 10^{49} \rm \, erg \, s$, $B_s = 2.525 \times 10^6 \rm \, G$, and $\delta = 3.864$. Then Figure 3 is a magnification of the graph between points $b$ and $c$ of Figure 2 together with the corresponding graph of the function $P_{RR} = P_{RR}(L)$ (i.e., the period of the perpendicular rotator represented by the rigidly rotating terminal model). The arrow in this figure represents the ``optimal nonaxisymmetric transition'' from the axisymmetric differentially rotating starting model (aligned rotator) with $L_{xx} = 2.154 \times 10^{49} \rm \, erg \, s$ and $P_{xx} = 24.949 \rm \, s$ to the nonaxisymmetric rigidly rotating terminal model (perpendicular rotator) with $L_{RR} = L_{xx}$ and $P_{RR} = 40.096 \rm \, s$. 

As discussed in \S \, 3, $P_{RR} = 2 \pi / \Omega_1$, where $\Omega_1$ is given in turn by equation (\ref{eq:LRR}) with the clarification that $I_{RR,11}(h,L) = A_r(I_1,L)I_{11}(h)$ (note that, for the needs of the present section, we shall write explicitly the angular momentum $L$ as second independent variable in this functional relation). The correction $A_r(I_1,L)$, so-called rigid rotation amplification ratio, is calculated by the algorithm developed in \S \, 5. Figure 4 shows the ``noncorrected value'' (``noncorrected'' in the sense that we attempt to approximate by this value the corresponding value of the perpendicular rotator) $I_{11}(h) = 1.323 \times 10^{50} \rm \, g \, cm^2$ and the corresponding ``corrected values'' $A_r(I_1,L)I_{11}(h)$ for several angular momentum values lying in the interval $[L_b,L_{x}]$. The arrow in this figure shows the corrected value for $L = L_{xx}$, $A_r(I_1,L_{xx})I_{11}(h) = 1.375 \times 10^{50} \rm \, g \, cm^2$; i.e., $A_r(I_1,L_{xx}) = 1.039$, which is equivalent to a percent correction $\sim 4 \%$ relative to $I_{11}(h)$. Now, as discussed in \S \, 4, the leading error involved in such calculations is due to the approximating relation (\ref{eq:IRRii}). Figure 5 gives $I_{RR,11}(h,L)$ as calculated (1) by 2DCIT, and (2) by the approximating relation (\ref{eq:IRRii}). This figure reveals a maximum error $\sim 0.25 \%$ relative to the 2DCIT value, which means in turn that the leading error involved in the calculation of the correction $A_r(I_1,L)$ is $\sim 20$ times less than the (percent) correction itself; consequently, $A_r(I_1,L)$ is a reliable estimate.             

Taking into account (1) the fundamental assumption of TOV scenario that the spin-down time rate remains constant during the turn-over and equal to the value observed now, 
\begin{equation} \label {eq:faTOV}
\dot{P} = {\rm constant} = \dot{P}_{now} = 5.64 \times 10^{-14} {\rm \, s \, s^{-1} \, ,}
\end{equation}
and (2) the above given optimal values (for $L_{xx}$, $B_s$, and $\delta$), we find $t_{DRD}= t_{TOV} = 8.517 \times 10^6 \rm \, yr$, $D\!E_{DRD} = D\!E_{TOV} = 1.008 \times 10^{47} \rm \, erg$, $t_{now} = (P_{now} - P_{xx}) / \dot{P}_{now} = 4.572 \times 10^6 \rm \, yr$, and $D_{TOV}(t_{now}) = D\!E_{DRD}/t_{DRD} = 3.754 \times 10^{32} \rm \, erg \, s^{-1}$; this value is $\sim 30$ times less than the classicaly estimated spin-down power (eq. [\ref{eq:Tdot}]) and thus the spin-down problem can be drastically simplified. 

Furthermore, assuming as above that secular evolution is initiated at the point $d$ of Fig. 2 and proceeds left to it, the following estimate for the secular timescale, $t_{SEC}$, can be given
\begin{equation} \label{eq:tSEC}
t_{SEC} \simeq \frac{P_{xx} - P_{min}}
                    {\left< \dot{P}_{sec} \right>_{[L_{xx},L_d]}} {\rm \, ,}
\end{equation}
where the subscript $[L_{xx},L_d]$ denotes the interval of angular momenta over which the average of the secular spin-down time rate is calculated. In the latter calculation, the magnetic flux is assumed invariant, $f_{xx} = f_d$; so, when we know the flux at $L_{xx}$, $f_{xx} = B_s R_{av}^2(L_{xx})$, we can find the average surface poloidal field holding at $d$, $B_d = B_s R_{av}^2(L_{xx})/R_{av}^2(L_d)$ and then, by using equation (\ref{eq:BsPs}), the corresponding secular spin-down time rate $\dot{P}_d$. The same can be done for any angular momentum lying in the interval $[L_{xx},L_d]$. We thus find $\left<\dot{P}_{sec}\right>_{[L_{xx},L_d]} \simeq 6.1 \times 10^{-16} \rm \, s \, s^{-1}$ and $t_{SEC} \simeq 6.4 \times 10^8 \rm \, yr$; thus, the secular timescale $t_{SEC}$ results $\sim 2$ orders of magnitude greater than the turn-over timescale $t_{TOV}$.         

Next, Figure 6 gives the graph of the function $\chi = \chi(t)$ in the time interval $[0,t_{TOV}]$ where, obviously, the time $t = 0$ corresponds to the beginning of the turn-over; the arrow head in this figure shows the ``current value'' $\chi_{now} = \chi(t_{now}) \simeq 71.6 \arcdeg$. Furthermore, Figure 7 gives the graph of the function $F_r = F_r(t)$, with the arrow head pointing to the current value $F_{r[now]} = F_r(t_{now}) \simeq 0.32$. Likewise, Figure 8 gives the graph of the function $D_{TOV} = D_{TOV}(t)$ and the arrow head shows the current value $D_{TOV[now]} = D_{TOV}(t_{now}) \simeq 3.8 \times 10^{32} \rm \, erg \, s^{-1}$. 

At this point, it is worth remarking that the time dependent scheme adopted in \S \, 6 (eq. [\ref{eq:tdep}]) is an alternative way to (re)calculate some significant time dependent quantities. This can be done by assuming an ``approximate rigid body description'' of our model. In particular, the time dependent angular momentum components $L_1(t)$ and $L_3(t)$ along the principal axes $I_1$ and $I_3$, respectively, can be written as (see, e.g., \cite{gold72}, eq. [5.20] for the rigid rotation analog) 
\begin{equation} \label{eq:L1t}
L_1(t) = L_{xx} \sqrt{1 -\nu_{\chi}(t)^2} =
         C_L(t) I_{11}(t) \Omega_{1c}(t) \left< \omega \right>_s\!(t) {\rm \, ,}
\end{equation}
\begin{equation} \label{eq:L3t}
L_3(t) = L_{xx} \nu_{\chi}(t) =
         C_L(t) I_{33}(t) \Omega_{3c}(t) \left< \omega \right>_s\!(t) {\rm \, ,}
\end{equation}
where $C_L(t)$ is a correction due to differential rotation and $\left< \omega \right>_s\!(t)$ is the average over the coordinate $s$ of the differential rotation function $\omega$ at the time $t$. For $t = 0$ we have $L_1(0) = 0$ and
\begin{equation} \label{eq:L30}
L_3(0) = L_{xx} = C_L(0) 
                  I_{DR,33}(h,L_{xx}) \Omega_{3c}(0) \left< \omega \right>_s\!(0) 
                                                                       {\rm \, ,}
\end{equation}
where $\Omega_{3c}(t) = Q(\Omega(L_{xx}),t)$ ($Q$ is the symbol for the time dependent scheme [\ref{eq:tdep}]) and thus $\Omega_{3c}(0) = \Omega(L_{xx})$; also, $\left< \omega \right>_s\!(t) = \left< \omega(F_r(t)) \right>_s = \left< \omega(Q(F_{r[ini]},t) \right>_s$ and thus $\left< \omega \right>_s\!(0) = \left< \omega(F_{r[ini]}) \right>_s$. So, the correction $C_L(0)$ is given by 
\begin{equation} \label{eq:CL0}
C_L(0) = \frac{L_{xx}}{I_{DR,33}(h,L_{xx}) \Omega(L_{xx})
         \left< \omega(F_{r[ini]}) \right>_s} {\rm \, .}
\end{equation}
Furthermore, $C_L(t_{TOV})=1$ since at the time $t_{TOV}$ rigid rotation has been established throughout the star. For the intermediate configurations, we assume the time dependent scheme (\ref{eq:tdep}),
\begin{equation} \label{eq:CLt}
C_L(t) \simeq 1 + Q \biggl( \bigl[C_L(0) - 1 \bigr],t \biggr) {\rm \, .}
\end{equation}
Thus, for the angular velocity components we can write
\begin{equation} \label{eq:W1ct}
\Omega_{1c}(t) = \frac{L_{xx} \sqrt{1 -\nu_{\chi}(t)^2}}
                      {C_L(t) I_{11}(t) \left< \omega \right>_s\!(t)} {\rm \, ,}
\end{equation}
\begin{equation} \label{eq:W3ct}
\Omega_{3c}(t) = \frac{L_{xx} \nu_{\chi}(t)}
                      {C_L(t) I_{33}(t) \left< \omega \right>_s\!(t)} {\rm \, .}
\end{equation}
Consequently, the resultant angular velocity along the $z$ axis (keep in mind, however, that the resultant angular velocity inclines at a small angle $\gamma$ with the $z$ axis), $\Omega_{zc}(t)$, is given by  
\begin{equation} \label{eq:Wzt}
\Omega_{zc}(t) = \sqrt{\Omega_{1c}^2(t) + \Omega_{3c}^2(t)} {\rm \, .}
\end{equation}  
The above involved time dependent moments of inertia $I_{11}(t)$ and $I_{33}(t)$ are approximated by the relations  
\begin{equation} \label{eq:I11t}
I_{11}(t) \simeq  
                A_r(I_1,L_{xx})I_{11}(h) - 
                Q \biggl( 
                \bigl[A_r(I_1,L_{xx})I_{11}(h) - I_{DR,11}(h,L_{xx})\bigr],t \biggr)
          {\rm \, ,}
\end{equation}
\begin{equation} \label{eq:I33t}
I_{33}(t) \simeq  
                I_{33}(h) + 
                Q \biggl( \bigl[I_{DR,33}(h,L_{xx}) - I_{33}(h)\bigr],t \biggr)
          {\rm \, .}
\end{equation}
For $t = 0$ we get $I_{11}(0) = I_{DR,11}(h,L_{xx})$ and $I_{33}(0) = I_{DR,33}(h,L_{xx})$; while for $t = t_{TOV}$ we find $I_{11}(t_{TOV}) = A_r(I_1,L_{xx})I_{11}(h)$ and $I_{33}(t_{TOV}) = I_{33}(h)$.

For the rotational kinetic energy $T(t)$ we can write (see, e.g., \cite{gold72}, eq. [5.21] for the rigid rotation analog)
\begin{equation} \label{eq:Tt}
T(t) = \frac{1}{2} C_T(t) \Biggl[ 
      I_{11}(t) \biggl( \Omega_{1c}(t) \left< \omega \right>_s\!(t) \biggr)^2 +
      I_{33}(t) \biggl( \Omega_{1c}(t) \left< \omega \right>_s\!(t) \biggr)^2
                        \Biggr] {\rm \, ,}
\end{equation}
where $C_T(t)$ is a correction due to differential rotation. For $t = 0$ we have 
\begin{equation}
T(0) = T_{xx} = \frac{1}{2} \, C_T(0) I_{DR,33}(h,L_{xx})
                \biggl( \Omega_{3c}(0) \left< \omega \right>_s\!(0) \biggr)^2 {\rm \, ,}
\end{equation}
which gives   
\begin{equation} \label{eq:CT}
C_T(0) = \frac{2 \, T_{xx}}{I_{DR,33}(h,L_{xx}) \left( 
      \Omega(L_{xx}) \left< \omega(F_{r[ini]}) \right>_s 
                                          \right)^2} {\rm \, .}
\end{equation}
Furthermore, $C_T(t_{TOV})=1$ since at the time $t_{TOV}$ rigid rotation has been established throughout the star. For the intermediate configurations, we adopt the time dependent scheme (\ref{eq:tdep}),
\begin{equation} \label{eq:CTt}
C_T(t) \simeq 1 + Q \biggl( \bigl[C_T(0) - 1 \bigr],t \biggr) {\rm \, .}
\end{equation}

Now, we are ready to (re)calculate some significant quantities on the basis of the approximate relations (\ref{eq:L1t})--(\ref{eq:CTt}). In particular, the central period $P(t)$ is written as
\begin{equation} \label{eq:Pt}
P(t)= \frac{2 \pi}{\Omega_{zc}(t)} {\rm \, ,}
\end{equation}
where $\Omega_{zc}(t)$ is given by equation (\ref{eq:Wzt}). On the other hand, however, the fundamental assumption (\ref{eq:faTOV}) leads to the linear relation 
\begin{equation} \label{eq:Ptlin}
P(t) = P_{RR} - (P_{RR} - P_{xx}) \nu_L(t) {\rm \, ,}
\end{equation}
where $\nu_L(t)$ is defined by equation (\ref{eq:nuLt}). To distinguish between the  values (\ref{eq:Pt})--(\ref{eq:Ptlin}) of $P(t)$, we add the subscript $FB$ to the first one, $P_{FB} = 2 \pi / \Omega_{zc}$, to give emphasis on that this is a ``feed-back'' value for $P(t)$. In order for the model to retain its inherent consistency, these two values should be close to each other, $(P(t)-P_{FB}(t)) \times 10^2 /P(t) < tol$, where $tol$ is a reasonable percent tolerance. Figure 9 shows (1) the function $P(t)$, which results from the assumption (\ref{eq:faTOV}), and (2) the function $P_{FB}(t)$ of which the time derivative $\dot{P}_{FB}$ is a feed-back value for the spin-down time rate, slightly deviating from the initial assumption (\ref{eq:faTOV}). This figure reveals a maximum percent difference $\sim 2 \%$ relative to $P(t)$, which, however, when taking into account the approximate character of equations (\ref{eq:L1t})--(\ref{eq:CTt})), seems to be insignificant. Furthermore, the corresponding average $\left< \dot{P}_{FB} \right>_t \simeq 5.6 \times 10^{-14} \rm \, s \, s^{-1}$ over the whole turn-over time deviates less than $\sim 1 \%$ from $\dot{P}_{now}$. In addition, Figure 10 shows (1) the central angular velocity $\Omega_{zc}(t)$ calculated with constant $\dot{P}$; (2) its counterpart $\Omega_{zc[FB]}$ calculated with feed-back $\dot{P}$; (3) the central angular velocity $\Omega_{3c}(t)$ calculated with constant $\dot{P}$; (4) its counterpart $\Omega_{3c[FB]}$ calculated with feed-back $\dot{P}$; (5) the central angular velocity $\Omega_{1c}(t)$ calculated with feed-back $\dot{P}$; and (6) the angular velocity $\Omega_z(t)$ of the progressively established rigid rotation in the mixture configuration approximated by the relation $\Omega_z(t) \simeq (\Omega_{zc}(t) - \Omega_{3c}(t))\left< \omega \right>_s \! (t)$.

Next, Figure 11 shows the rotational kinetic energy $T(t)$ of the model, calculated (1) by the relation $T(t) = T_{xx} - \int_0^{t}{D_{TOV}(t)} \, dt$ (where $D_{TOV}(t)$ is given by eq. [\ref{eq:tdottov}] with $t_{TOV}$ substituted by $t$), and (2) by the relation (\ref{eq:Tt}) which, as discussed above, represents the corresponding feed-back value $T_{FB}(t)$. As seen in this figure, the extremum percent difference is $\sim -1.8 \%$ relative to $T(t)$ near $t = 4 \times 10^6 \rm \, yr$. Since, however, in our model the quantity $\dot{T}$ is of higher significance than $T$ itself, it is interesting to remark that these two graphs exhibit almost equal time rates; in addition, their average time rates over the whole turn-over time are almost equal: $\left<\dot{T}_{FB}\right>_t \simeq \left<\dot{T}\right>_t = - 3.754 \times 10^{32} \rm \, erg \, s^{-1}$. 

Figure 12 shows the angle $\gamma(t)$ between the instantaneous angular velocity axis and the invariant angular momentum axis. As it can be verified after some elementary algebra, this angle is given by 
\begin{equation} \label{eq:gamma}
\gamma(t) = \frac{L_{xx}}{\Omega_{zc}(t)}
            \left(
                  \frac{1 - \nu_{\chi}^2(t)}{I_{11}(t)} +
                  \frac{\nu_{\chi}^2(t)}{I_{33}(t)}
            \right) {\rm \, .}
\end{equation}
The arrow head in this figure shows the current value $\gamma_{now} = \gamma(t_{now}) \simeq 3100 \rm \, arcseconds$. The maximum value of $\gamma$ is $\sim 3500 \rm \, arcseconds$ near $t = 3.5 \times 10^6 \rm \, yr$; thus, as emphatically suggested in the foregoing sections, the angle $\gamma$ remains less than $1 \arcdeg$ during the whole turn-over.

Finally, Table 1 gives a summary of calculations for both the cases $F_{r[ini]} = 1.00 {\rm \ and \ } 0.50$. Apparently, for the latter value of $F_r$ the ratio $\frac{\dot{T}({\rm of \, eq. \, [\ref{eq:Tdot}]})}{\left< \dot{T} \right>_t} \sim 50$ leads to further simplification of the spin-down problem in AE Aquarii. Concluding, we find interesting to emphasize once again on the remark made in the Abstract: ``an observed large spin-down time rate does not always imply a large spin-down power''.   

\acknowledgments{
The TOV scenario has resulted as a product (contributed by the author) within the framework of a research project supported by the Research Commitee of the University of Patras (C. Carathe\'odory's Research Project 1998/1932). It is a pleasure for the author to acknowledge his colleagues in this project Dr. Florentia Valvi and Dr. Panagiotis Papasotiriou for helpful discussions and comments on the manuscript.}  

\clearpage

\begin{deluxetable}{lrr}
\scriptsize
\tablecolumns{3}
\tablewidth{0pt}
\tablecaption{Summary of calculations regarding TOV 
              scenario\tablenotemark{a}  \label{tbl-1}}
\tablehead{
\colhead{} & \multicolumn{2}{c}{$F_{r[ini]}$} \\
\cline{2-3}                           \\
\colhead{Parameter\tablenotemark{b}} 
& \colhead{1.00}  
& \colhead{0.50}
}
\startdata
Optimal $L_{xx}$                          & 2.154$(+49)$\tablenotemark{c} & 2.272$(+49)$
\nl
Optimal $B_s$                                              & 2.525$(+06)$ & 1.018$(+06)$
\nl
Optimal $\delta$ (dimensionless)                           & 3.864$(+00)$ & 7.160$(+00)$
\nl
$\left< \sigma_{\perp}(H_t) / \sigma_c \right>$ \tablenotemark{d}
                                                           & 1.105$(+03)$ & 1.001$(+03)$
\nl               
Magnetic flux, $f$                                         & 1.149$(+24)$ & 4.663$(+23)$
\nl
Rigid rotation amplification ratio, $A_r(I_1,L_{xx})$ (dimensionless)  
                                                           & 1.039$(+00)$ & 1.043$(+00)$
\nl
Secular timescale, $t_{SEC}$ (yr)                          & 6.392$(+08)$ & 7.621$(+09)$
\nl
TOV \& DRD timescales, $t_{TOV}=t_{DRD}$ (yr)              & 8.517$(+06)$ & 4.873$(+06)$
\nl
TOV \& DRD timescales in units of $P_{xx}$ (number of revolutions)   
                                                           & 1.077$(+13)$ & 5.208$(+12)$
\nl
Present TOV time, $t_{now}$ (yr)                           & 4.572$(+06)$ & 2.009$(+06)$
\nl
Present spin-down power, $D_{TOV}(t_{now})=\frac{D\!E_{DRD}}{t_{DRD}} 
                \equiv \left< \dot{T} \right>_t$ \tablenotemark{e} 
                                                           & 3.754$(+32)$ & 2.218$(+32)$
\nl
Ratio $\frac{\dot{T}({\rm of \, eq. \, [\ref{eq:Tdot}]})}{\left< \dot{T} \right>_t}$ 
                                                           & 2.664$(+01)$ & 4.509$(+01)$
\nl
Power dissipated due to electron and ion viscosity $\mu_{e+i}$, 
                                    $D_{EIV}$ \tablenotemark{f}
                                                           & 3.467$(+30)$ & 1.201$(+30)$
\nl 
Present turn-over angle, $\chi_{now}$ (arcdegrees)         & 7.160$(+01)$ & 7.071$(+01)$
\nl
Present reduction factor, $F_{r[now]}$ (dimensionless)     & 3.156$(-01)$ & 1.652$(-01)$
\nl
Present central angular velocity about the $I_3$ axis, $\Omega_{3c[now]}$
                                                           & 7.849$(-02)$ & 7.036$(-02)$
\nl
Present angular velocity -- angular momentum angle, $\gamma_{now}$ (arcseconds)
                                                           & 3.102$(+03)$ & 3.294$(+03)$
\nl
\cutinhead{Calculations for the axisymmetric differentially rotating starting model 
           (aligned rotator)}
Central period, $P_{xx}$                                   & 2.495$(+01)$ & 2.951$(+01)$
\nl
Rotational kinetic energy, $T_{xx}$                        & 1.789$(+48)$ & 1.904$(+48)$
\nl
Moment of inertia along the $I_1$ axis, $I_{DR,11}(h,L_{xx})$
                                                           & 1.387$(+50)$ & 1.391$(+50)$
\nl
Moment of inertia along the $I_3$ axis, $I_{DR,33}(h,L_{xx})$
                                                           & 1.376$(+50)$ & 1.384$(+50)$
\nl
Average radius, $R_{av}$                                   & 6.744$(+08)$ & 6.767$(+08)$
\nl
Radius (spherical approximation), $R_{sph}$                & 6.525$(+08)$ & 6.527$(+08)$
\nl
Volume (spherical approximation), $V_{sph}$                & 1.164$(+27)$ & 1.165$(+27)$
\nl
Average density (spherical approximation), $\left< \varrho_{sph} \right>$
                                                           & 1.490$(+06)$ & 1.487$(+06)$
\nl
Average pressure (spherical approximation), $\left< P_{sph} \right>$
                                                           & 4.450$(+23)$ & 4.439$(+23)$
\nl
Average sound speed (spherical approximation), $\left< a_{sph} \right>$
                                                           & 5.466$(+08)$ & 5.463$(+08)$
\nl
Moments of inertia (spherical approximation), $I_{11[sph]}=I_{22[sph]}=I_{33[sph]}$
                                                           & 1.285$(+50)$ & 1.286$(+50)$
\nl
Dimensionless radius (spherical approximation), $\xi_1$    & 3.711$(+00)$ & 3.710$(+00)$
\nl
Dimensionless base of the surface zone (spherical approximation), $\xi_s$
                                                           & 3.628$(+00)$ & 3.627$(+00)$
\nl
Average surface poloidal magnetic field, $B_s$             & 2.525$(+06)$ & 1.018$(+06)$
\nl
Proper poloidal magnetic parameter, $\beta_{*}^p$          & 2.011$(+04)$ & 4.985$(+04)$
\nl
Average surface toroidal magnetic field, $\left< H_t(\xi_s,\nu) \right>_\nu$
                                                           & 2.810$(+09)$ & 2.811$(+09)$ 
\nl
Surface Alfv\'en speed (spherical approximation), $V_{As[sph]}$
                                                           & 6.501$(+03)$ & 2.621$(+03)$ 
\nl
Surface Alfv\'en time (spherical approximation), $t_{As[sph]}=R_{sph}/V_{As[sph]}$
                                                           & 1.004$(+05)$ & 2.490$(+05)$
\nl 
Average Alfv\'en speed (spherical approximation), 
             $V_{A[sph]}=\frac{<H_p>}{\sqrt{4\pi <\varrho_{sph}>}}$
                                                           & 8.618$(+03)$ & 3.475$(+03)$
\nl
Average Alfv\'en time (spherical approximation), $t_{A[sph]}=R_{sph}/V_{A[sph]}$
                                                           & 7.571$(+04)$ & 1.879$(+05)$
\nl
Maximum poloidal magnetic field, $H_{p[max]}$              & 1.172$(+08)$ & 4.722$(+07)$
\nl
Average poloidal magnetic field, $\left< H_p \right>$                   & 3.729$(+07)$ & 1.502$(+07)$
\nl
Average $z$-component of the poloidal magnetic field, $\left< H_{pz} \right>$
                                                           & 3.461$(+07)$ & 1.394$(+07)$
\nl
Average $\varpi$-component of the poloidal magnetic field, $\left< H_{p\varpi} \right>$
                                                           & 7.271$(+06)$ & 2.929$(+06)$
\nl
Maximum toroidal magnetic field, $H_{t[max]}$              & 7.760$(+11)$ & 7.750$(+11)$ 
\nl
Average toroidal magnetic field, $\left< H_t \right>$      & 3.198$(+11)$ & 3.194$(+11)$
\nl
Ratio $\Omega_c / \Omega_e = 1 / \omega_e$                 & 3.625$(+00)$ & 2.125$(+00)$
\nl
\cutinhead{Calculations for the nonaxisymmetric rigidly rotating terminal model 
           (perpendicular rotator)}
Period, $P_{RR}$                                           & 4.010$(+01)$ & 3.817$(+01)$
\nl
Rotational kinetic energy, $T_{RR}$                        & 1.688$(+48)$ & 1.870$(+48)$
\nl
Moment of inertia along the $I_1$ axis, $I_{RR,11}(h,L_{xx})$
                                                           & 1.375$(+50)$ & 1.380$(+50)$
\nl
Moment of inertia along the $I_3$ axis, $I_{33}(h)$
                                                           & 1.283$(+50)$ & 1.283$(+50)$
\nl
Average radius, $R_{av}$                                   & 6.574$(+08)$ & 6.574$(+08)$
\nl
Radius (spherical approximation), $R_{sph}$                & 6.409$(+08)$ & 6.409$(+08)$
\nl
Volume (spherical approximation), $V_{sph}$                & 1.103$(+27)$ & 1.103$(+27)$
\nl
Average density (spherical approximation), $\left< \varrho_{sph} \right>$
                                                           & 1.600$(+06)$ & 1.600$(+06)$
\nl
Average pressure (spherical approximation), $\left< P_{sph} \right>$
                                                           & 5.012$(+23)$ & 5.012$(+23)$
\nl
Average sound speed (spherical approximation), $\left< a_{sph} \right>$
                                                           & 5.598$(+08)$ & 5.598$(+08)$
\nl
Moments of inertia (spherical approximation), $I_{11[sph]}=I_{22[sph]}=I_{33[sph]}$
                                                           & 1.255$(+50)$ & 1.255$(+50)$
\nl
Dimensionless radius (spherical approximation), $\xi_1$    & 3.730$(+00)$ & 3.730$(+00)$
\nl
Dimensionless base of the surface zone (spherical approximation), $\xi_s$
                                                           & 3.650$(+00)$ & 3.650$(+00)$
\nl
Average surface poloidal magnetic field, $B_s$             & 2.658$(+06)$ & 1.079$(+06)$
\nl
Proper poloidal magnetic parameter, $\beta_{*}^p$          & 1.935$(+04)$ & 4.765$(+04)$
\nl
Average surface toroidal magnetic field, $\left< H_t(\xi_s,\nu) \right>_\nu$
                                                           & 2.762$(+09)$ & 2.762$(+09)$ 
\nl
Surface Alfv\'en speed (spherical approximation), $V_{As[sph]}$
                                                           & 6.842$(+03)$ & 2.778$(+03)$ 
\nl
Surface Alfv\'en time (spherical approximation), $t_{As[sph]}$
                                                           & 9.368$(+04)$ & 2.307$(+05)$
\nl 
Average Alfv\'en speed (spherical approximation), $V_{A[sph]}$
                                                           & 9.146$(+03)$ & 3.714$(+03)$
\nl
Average Alfv\'en time (spherical approximation), $t_{A[sph]}$
                                                           & 7.007$(+04)$ & 1.726$(+05)$
\nl
Maximum poloidal magnetic field, $H_{p[max]}$              & 1.299$(+08)$ & 5.274$(+07)$
\nl
Average poloidal magnetic field, $\left< H_p \right>$                   & 4.101$(+07)$ & 1.665$(+07)$
\nl
Average $z$-component of the poloidal magnetic field, $\left< H_{pz} \right>$
                                                           & 3.806$(+07)$ & 1.545$(+07)$
\nl
Average $\varpi$-component of the poloidal magnetic field, $\left< H_{p\varpi} \right>$
                                                           & 8.004$(+06)$ & 3.250$(+06)$
\nl
Maximum toroidal magnetic field, $H_{t[max]}$              & 8.252$(+11)$ & 8.252$(+11)$ 
\nl
Average toroidal magnetic field, $\left< H_t \right>$      & 3.387$(+11)$ & 3.387$(+11)$
\nl
Ratio $\Omega_c / \Omega_e = 1 / \omega_e$                 & 1.000$(+00)$ & 1.000$(+00)$
\enddata
\tablenotetext{a}{Basic model: 
                  $M = 0.89 \, M_{\sun}$; 
                  $h = 0.08$; 
                  $\kappa = \cos 45 \arcdeg$ = 0.7071; 
                  $\chi_{ini} = \pi / 128 \simeq 1.41 \arcdeg$; 
                  $\nu_{\chi[ini]} = 0.9997$} 
\tablenotetext{b}{In cgs units, unless stated otherwise}
\tablenotetext{c}{Parenthesized numbers denote powers of 10}
\tablenotetext{d}{Typical values for a carbon plasma with 
                  $\frak{L} = 10^{32} \rm \, erg \, s^{-1}$, 
                  $\frak{T}_s \simeq 3.97 \times 10^7 \rm \, K$,
                  $\varrho_s \simeq 1.2 \times 10^4 \rm \, g \, cm^{-3}$}
\tablenotetext{e}{Involved parameters: 
                  $A_P = 1$; 
                  $A_{TOV}^h = 10$; 
                  $A_{TOV}^v = 2 \times 10$ (for details, see \S \, 6)}
\tablenotetext{f}{Details as in \cite{duri73}, \S \, 3, especially eqs. (4), (6)} 
\end{deluxetable}

\clearpage

\figcaption[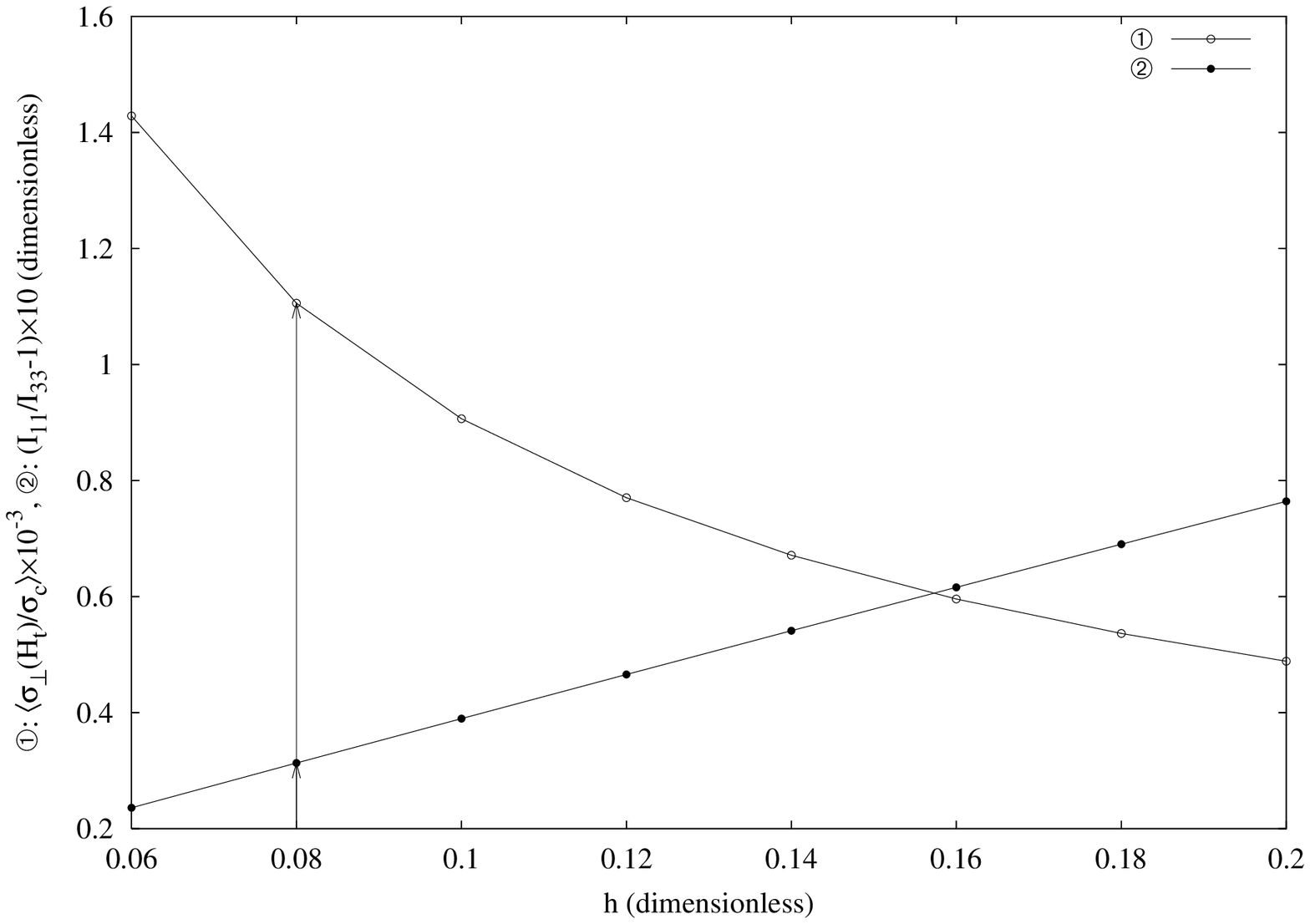]{The quantities $\left< \sigma_\perp(H_t) / \sigma_c \right> \times 10^{-3}$ and $(I_{11}/I_{33} -1) \times 10$ as functions of $h$. The double-head arrow has base at $h = 0.08$ and shows for the former and latter quantities their corresponding values $1.105$ and $0.313$, respectively. This is an ``optimal value'' for the magnetic perturbation parameter $h$ in the sense that at this value the former quantity is still greater than unity, while at the same time the latter quantity shows that dynamical asymmetry has been already established. \label{fig1}}

\figcaption[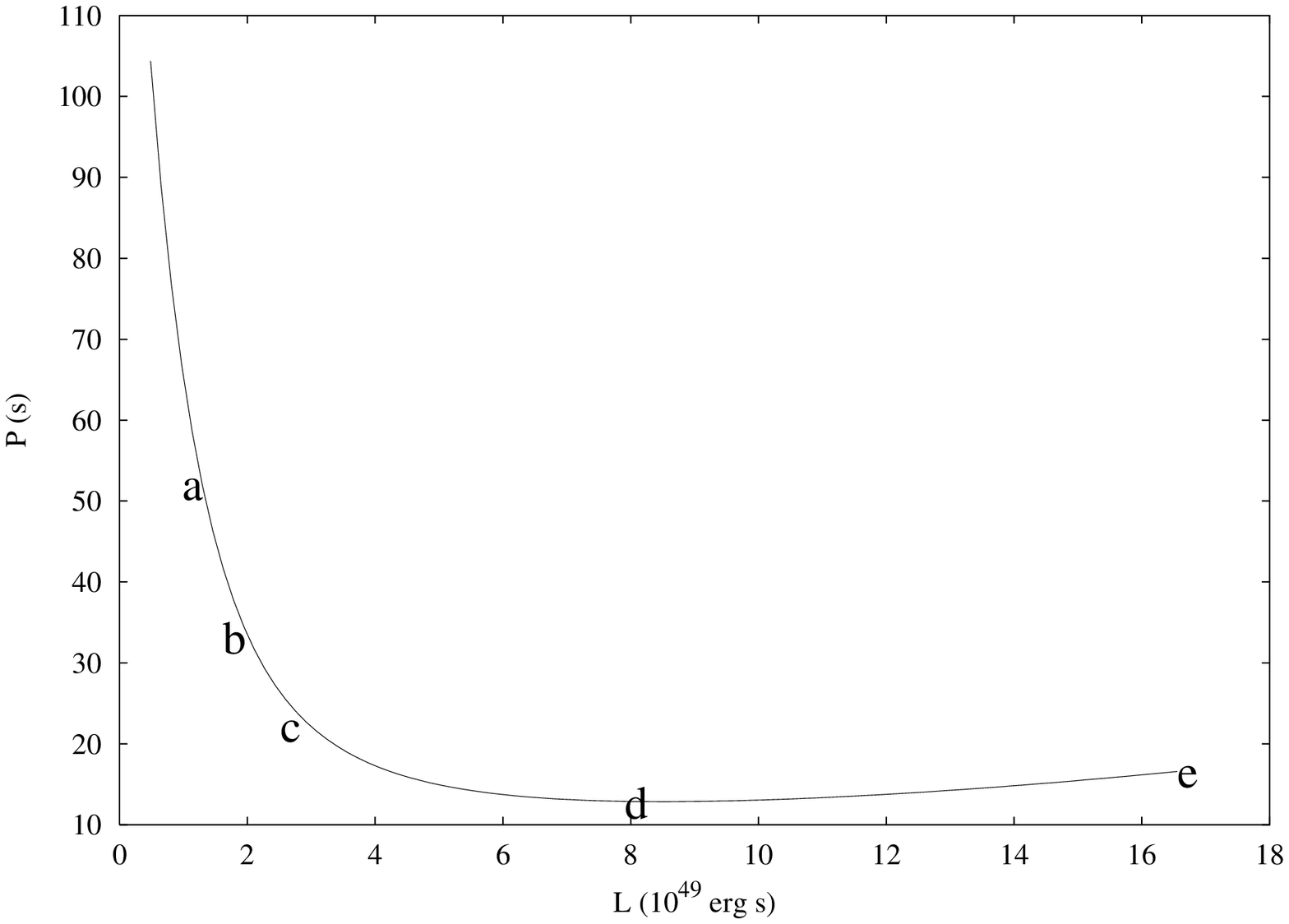]{$P$ vs. $L$ for the constant-mass sequence $[M = 0.89 M_{\sun}, h=0.08, F_r =1.00]$. Point $a(L=0.979 \times 10^{49} {\rm \, erg \, s}, P=52.091 {\rm \, s})$: almost spherical model due to counterbalancing of the effects of rotation and poloidal field with the effects of toroidal field. Point $b(L=1.615 \times 10^{49} {\rm \, erg \, s}, P=33.080 {\rm \, s})$: model with period equal to $P_{now}$; so, $L(P_{now}) = L_b = 1.615 \times 10^{49} {\rm \, erg \, s}$. Point $c(L=2.519 \times 10^{49} {\rm \, erg \, s},  P=22.150 {\rm \, s})$: model for which $I_{11} = I_{33}$ (dynamical asymmetry holds left to this point); so, $L_{x} = L_c = 2.519 \times 10^{49} \rm \, erg \, s$. Point $d(L=7.904 \times 10^{49} {\rm \, erg \, s}, P=12.634 {\rm \, s})$: model of minimum period, $P_{min} = 12.634 \rm \, s$; so, $L(P_{min}) = L_d = 7.904 \times 10^{49} \rm \, erg \, s$. Point $e(L=16.553 \times 10^{49} {\rm \, erg \, s}, P=16.582 {\rm \, s})$: critically rotating model with ratio of kinetic to gravitational energy $\tau = T / |W| = 0.140$ and effective ellipticity $e_{eff} = 0.388$. \label{fig2}}

\figcaption[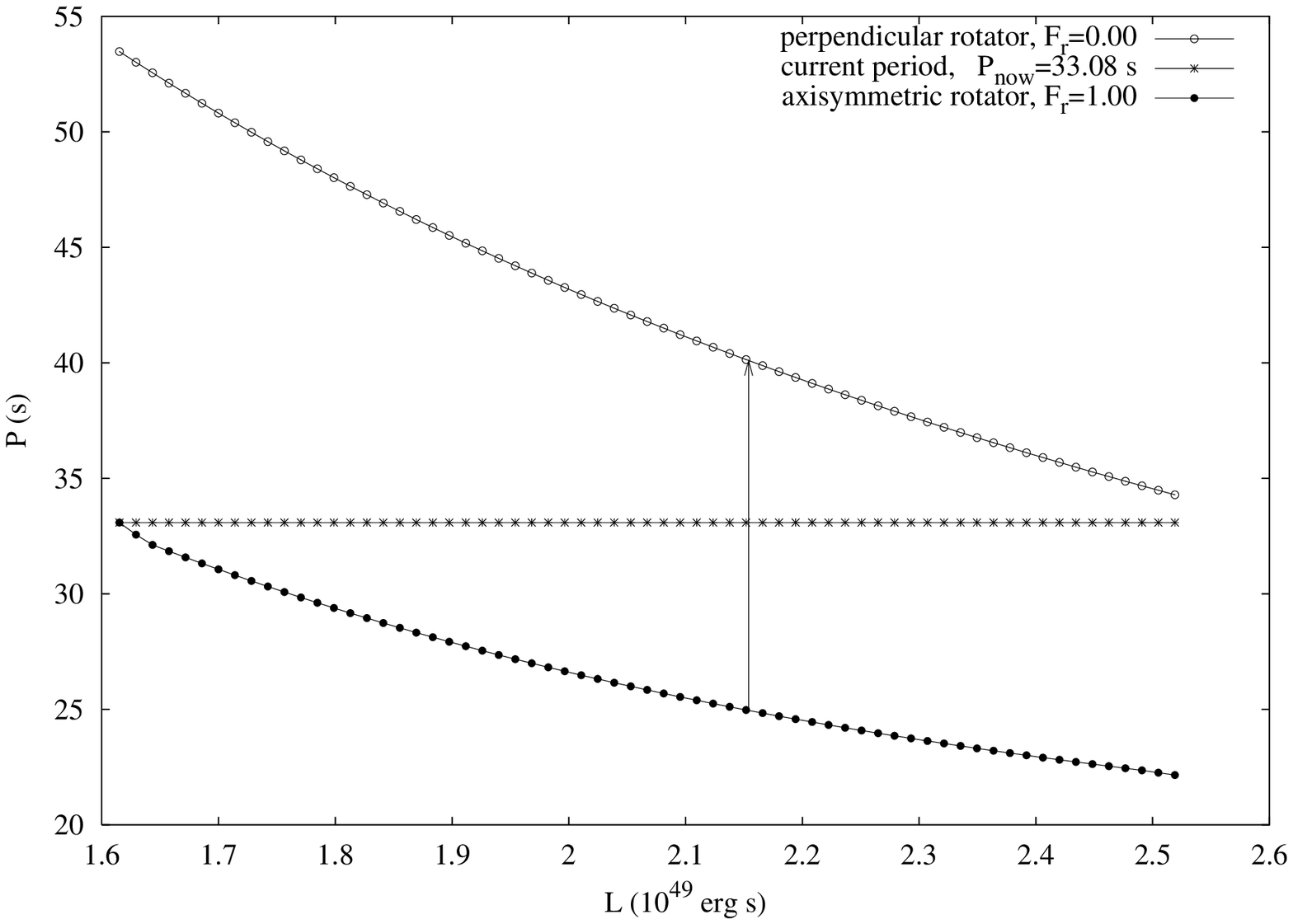]{Magnification of the graph between points $b$ and $c$ of Fig. 2 together with the corresponding graph of the function $P_{RR} = P_{RR}(L)$, i.e., the period of the perpendicular rotator represented by the rigidly rotating terminal model. The arrow shows an ``optimal nonaxisymmetric transition'' from the axisymmetric differentially rotating starting model (aligned rotator) with $L_{xx} = 2.154 \times 10^{49} \rm \, erg \, s$ and $P_{xx} = 24.949 \rm \, s$ to the nonaxisymmetric rigidly rotating terminal model (perpendicular rotator) with $L_{RR} = L_{xx}$ and $P_{RR} = 40.096 \rm \, s$. \label{fig3}}

\figcaption[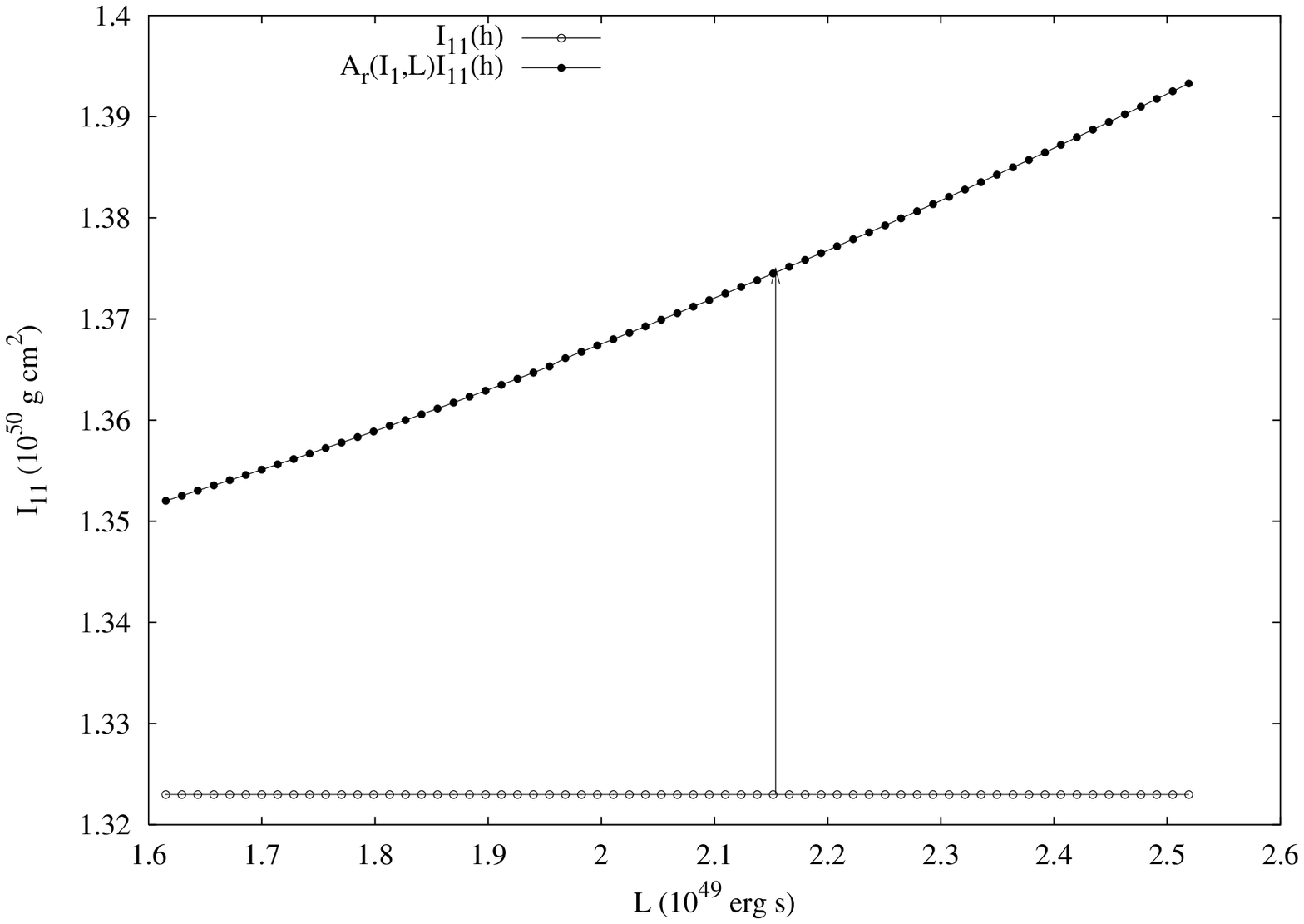]{The ``noncorrected value'' $I_{11}(h) = 1.323 \times 10^{50} \rm \, g \, cm^2$ (``noncorrected'' in the sense that we attempt to approximate by this value the corresponding value of the perpendicular rotator) and the corresponding ``corrected values'' $A_r(I_1,L)I_{11}(h)$ vs. $L$ in the interval $[L_b,L_{x}]$. The arrow shows the corrected value for $L = L_{xx}$, $A_r(I_1,L_{xx})I_{11}(h) = 1.375 \times 10^{50} \rm \, g \, cm^2$; i.e., $A_r(I_1,L_{xx}) = 1.039$, which is equivalent to a percent correction $\sim 4 \%$ relative to $I_{11}(h)$. \label{fig4}}

\figcaption[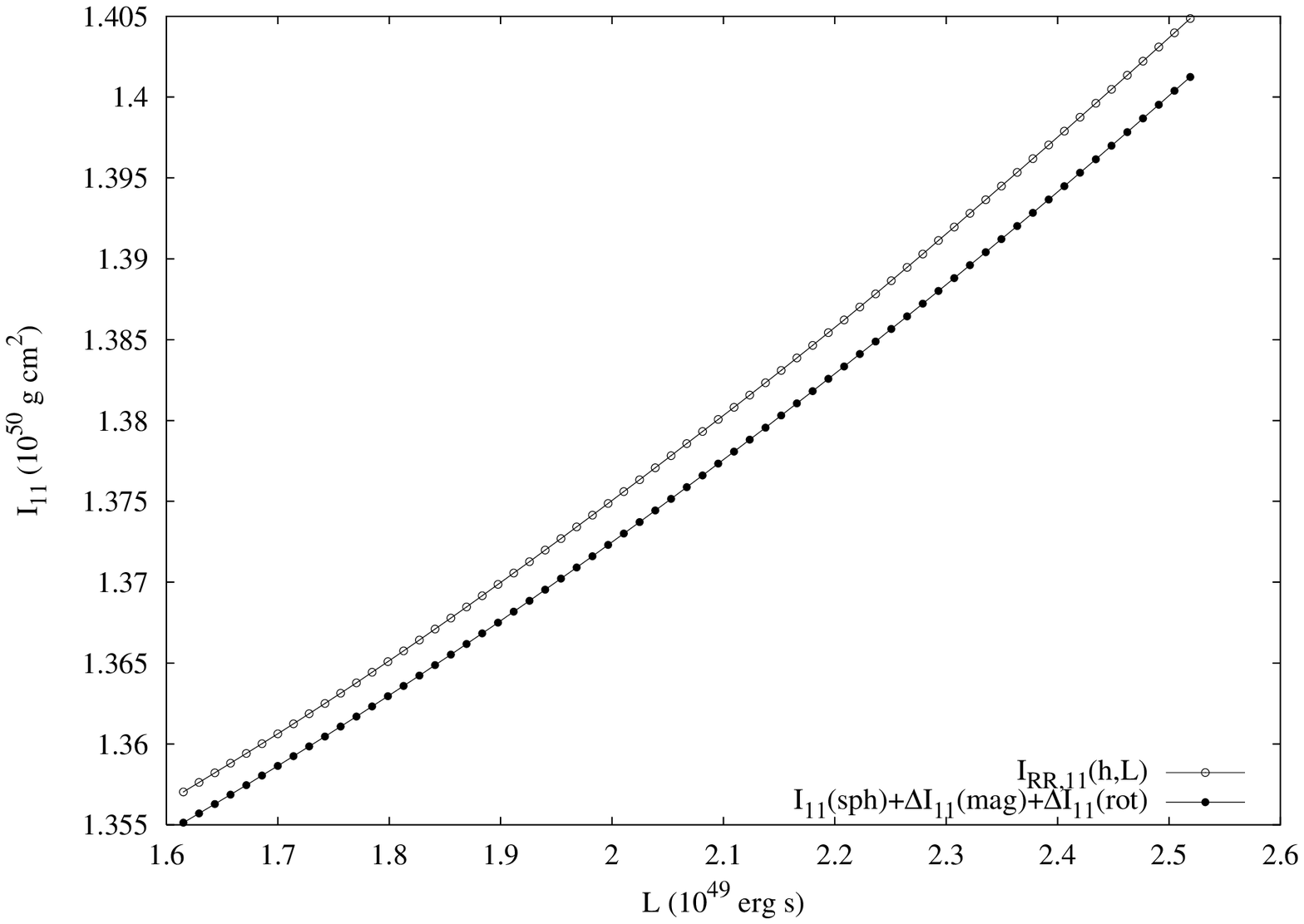]{$I_{RR,11}(h,L)$ calculated (1) by 2DCIT and (2) by the approximating relation (\ref{eq:IRRii}) vs. $L$ in the interval $[L_b,L_{x}]$. This figure reveals a maximum difference $\sim 0.25 \%$ of the latter relative to the former values, which means in turn that the leading error involved in the calculation of $A_r(I_1,L)$ is $\sim 20$ times less than the (percent) correction itself. \label{fig5}}

\figcaption[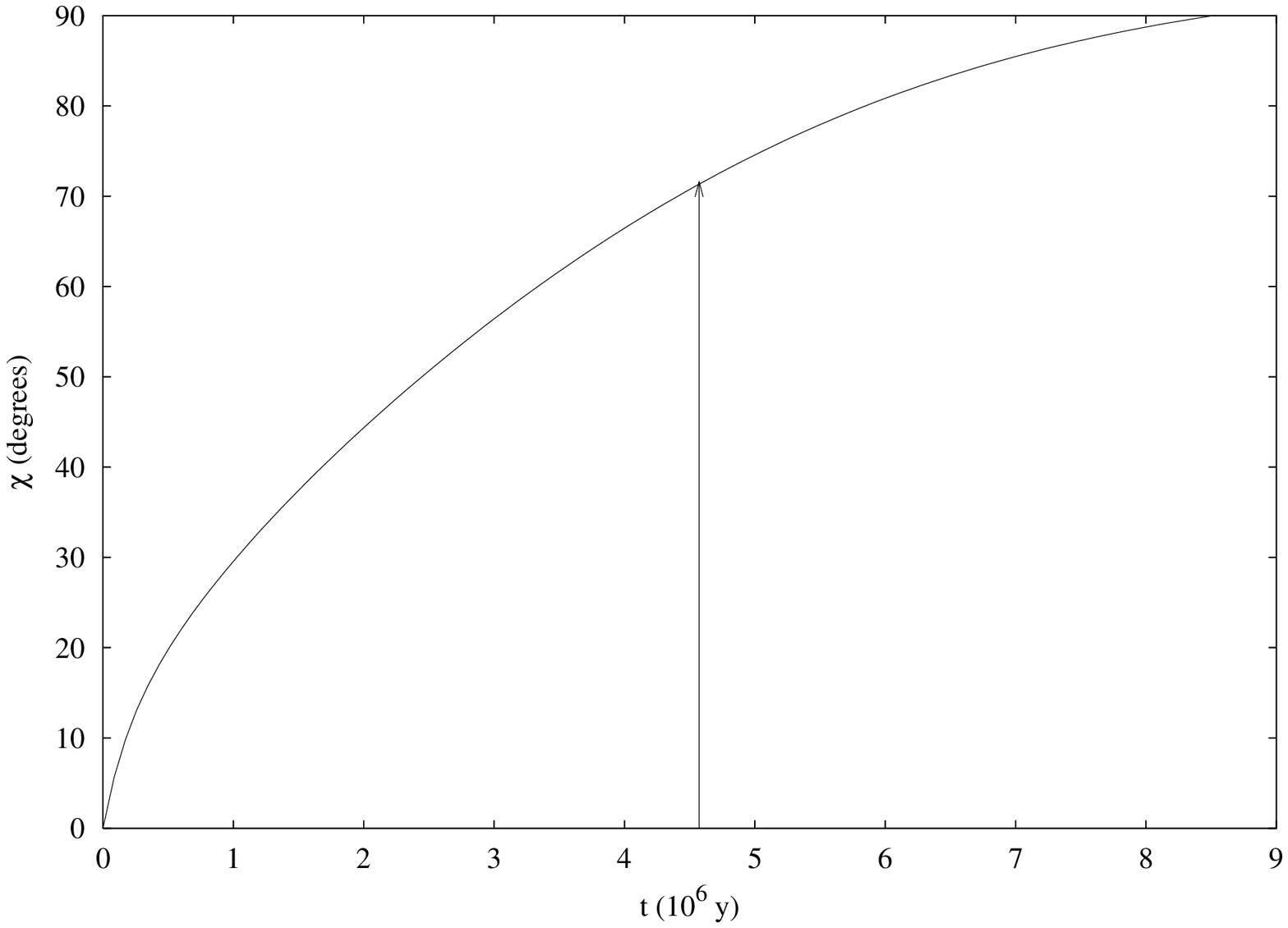]{The turn-over angle $\chi$ vs. $t$ in the interval $[0,t_{TOV}]$. The arrow head shows the ``current value'' $\chi_{now} = \chi(t_{now}) = 71.604 \arcdeg$. \label{fig6}}

\figcaption[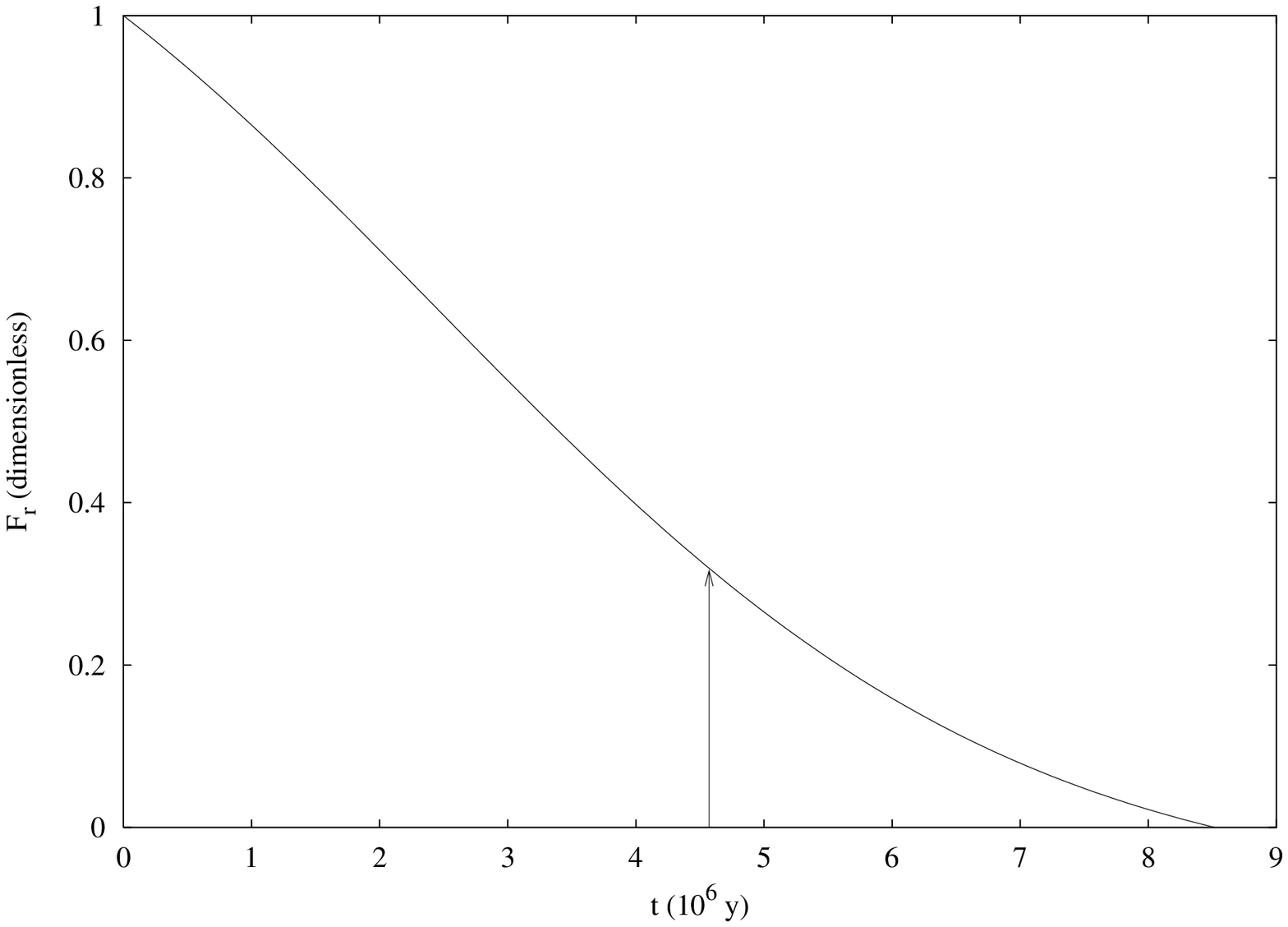]{The differential rotation reduction factor $F_r$ vs. $t$ in the interval $[0,t_{TOV}]$. The arrow head points to the current value $F_{r[now]} = F_r(t_{now}) = 0.316$. \label{fig7}}

\figcaption[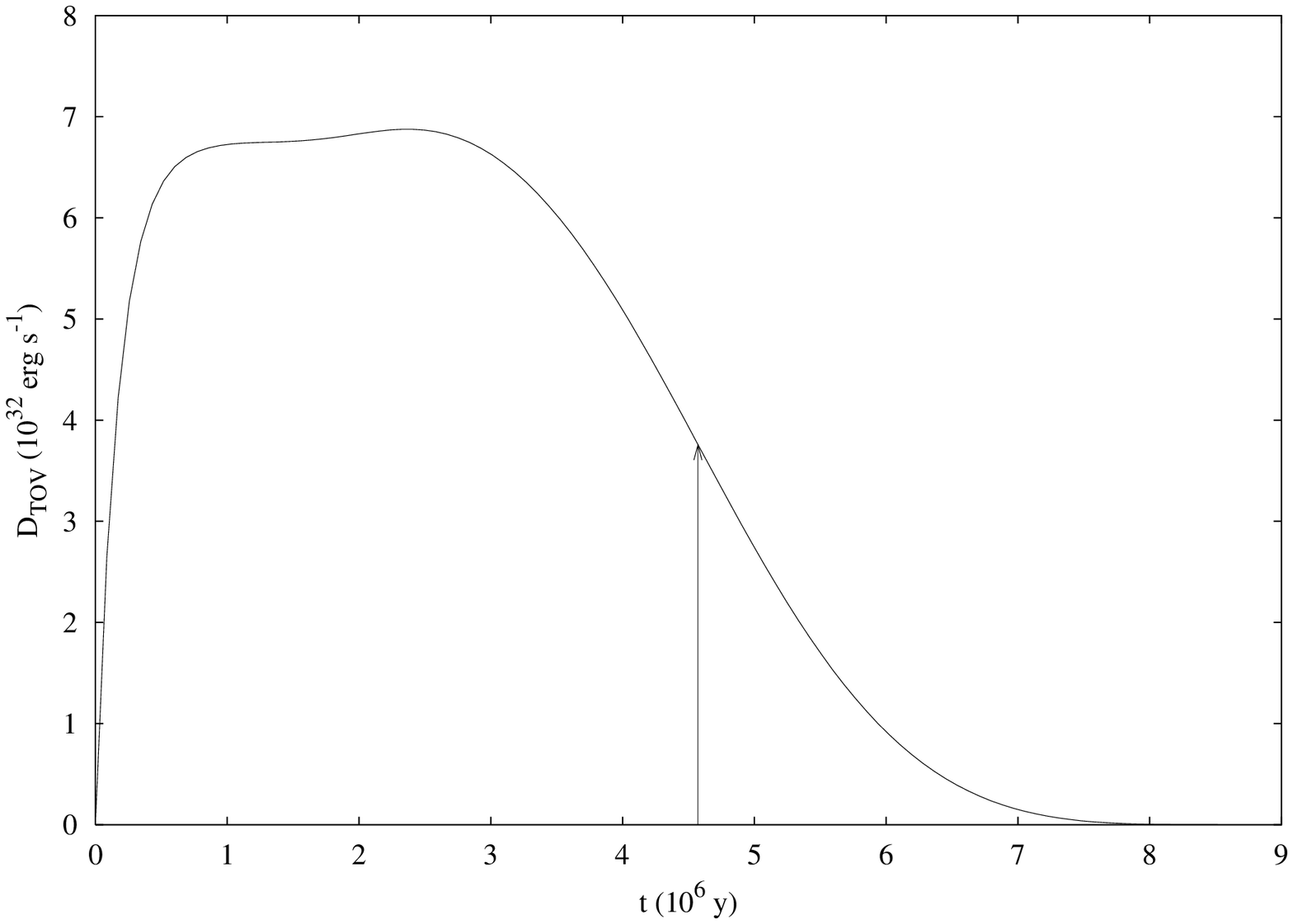]{The power $D_{TOV}$ dissipated due to turn-over vs. $t$ in the interval $[0,t_{TOV}]$. The arrow head shows the current value $D_{TOV[now]} = D_{TOV}(t_{now}) = 3.754 \times 10^{32} \rm \, erg \, s^{-1}$. \label{fig8}}

\figcaption[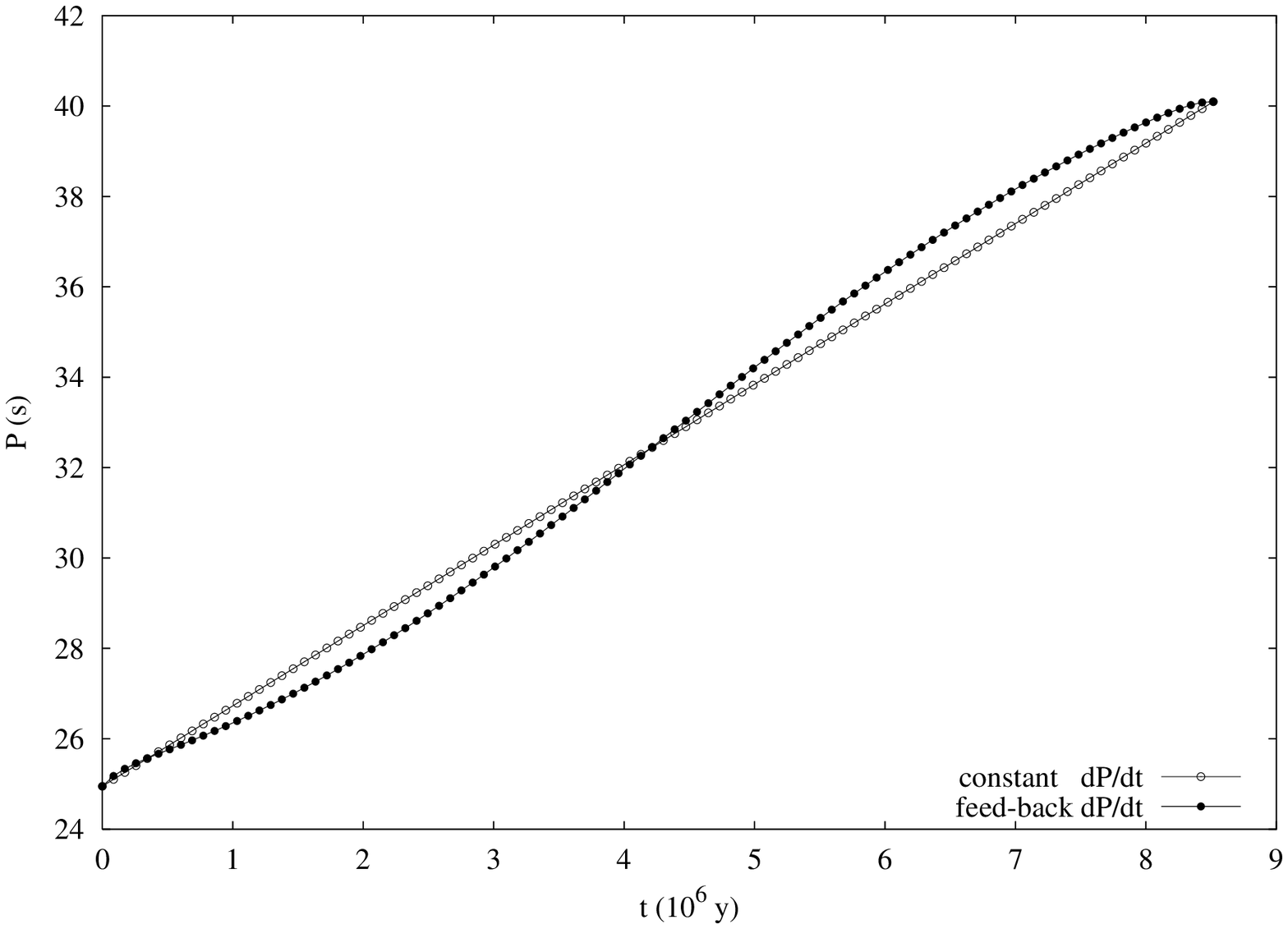]{The central period $P(t)$ as it results (1) from the assumption (\ref{eq:faTOV}) and (2) from the function $P_{FB}(t)$ (eq. [\ref{eq:Pt}]) of which the time derivative $\dot{P}_{FB}$ is a feed-back value for the spin-down time rate, slightly deviating from the initial assumption (\ref{eq:faTOV}). This figure reveals a maximum percent difference $\sim 2 \%$ of the latter relative to the former values; however, when taking into account the approximate character of equations (\ref{eq:L1t})--(\ref{eq:CTt}), this difference seems to be insignificant. \label{fig9}}

\figcaption[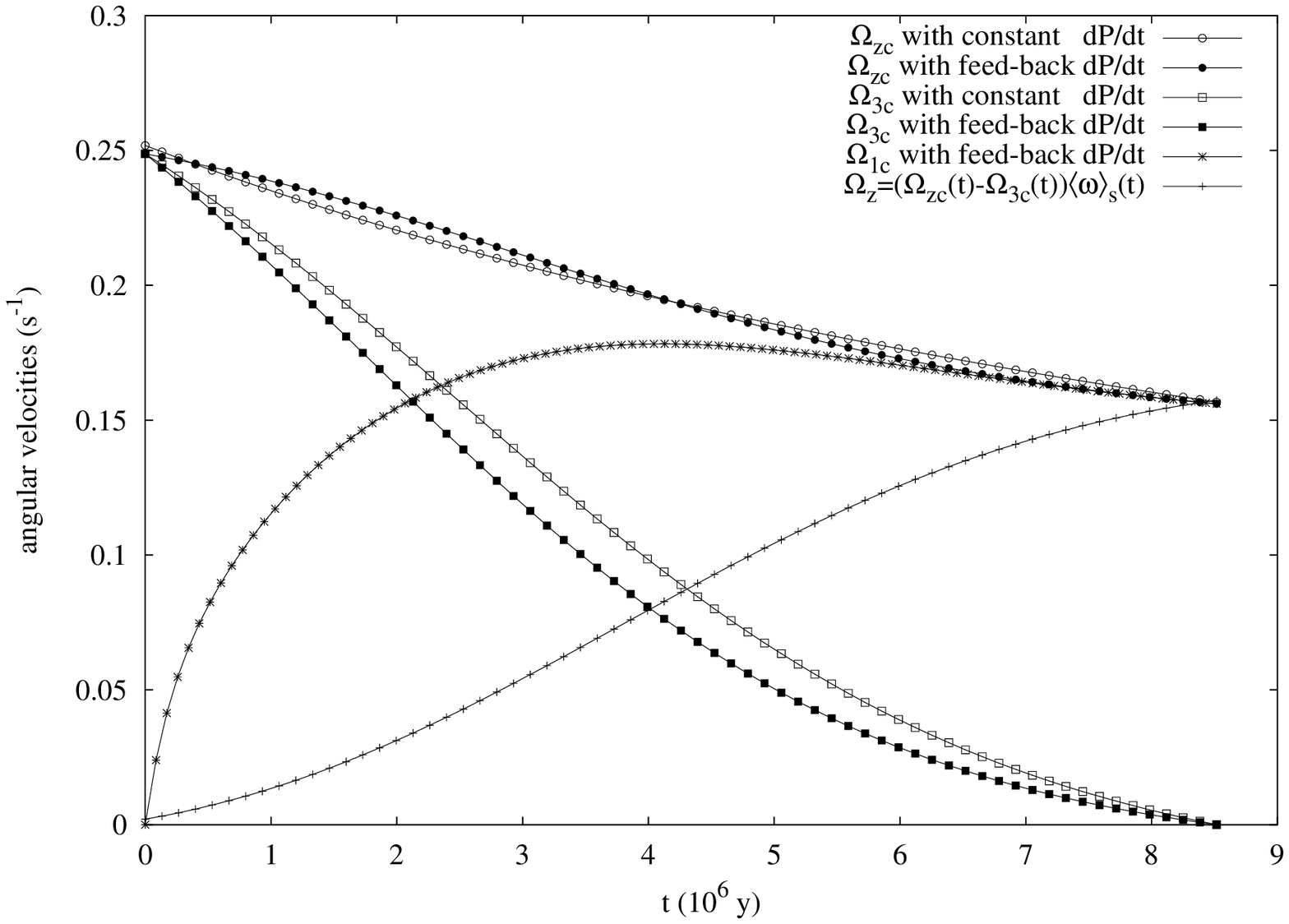]{(1) the central angular velocity $\Omega_{zc}(t)$ calculated with constant $\dot{P}$; (2) its counterpart $\Omega_{zc[FB]}$ calculated with feed-back $\dot{P}$; (3) the central angular velocity $\Omega_{3c}(t)$ calculated with constant $\dot{P}$; (4) its counterpart $\Omega_{3c[FB]}$ calculated with feed-back $\dot{P}$; (5) the central angular velocity $\Omega_{1c}(t)$ calculated with feed-back $\dot{P}$; and (6) the angular velocity $\Omega_z(t)$ of the progressively established rigid rotation in the mixture configuration approximated by the relation $\Omega_z(t) \simeq (\Omega_{zc}(t) - \Omega_{3c}(t))\left< \omega \right>_s \! (t)$. \label{fig10}}
 
\figcaption[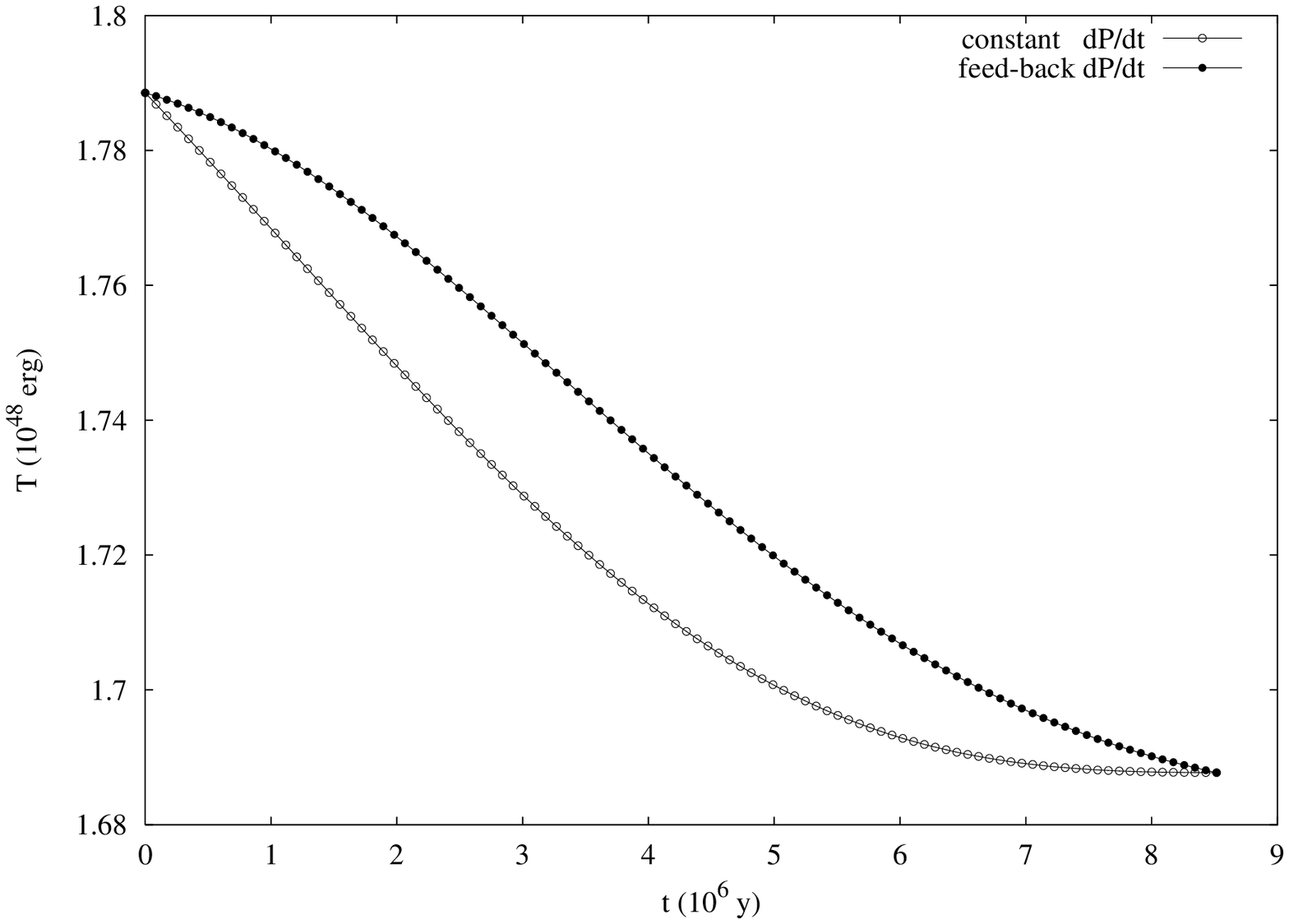]{The rotational kinetic energy $T$ calculated (1) by the relation $T(t) = T_{xx} - \int_0^{t}{D_{TOV}(t)} \, dt$ (where $D_{TOV}(t)$ is given by eq. [\ref{eq:tdottov}] with $t_{TOV}$ substituted by $t$), and (2) by the relation (\ref{eq:Tt}) which represents the corresponding feed-back value $T_{FB}(t)$. The extremum percent difference of the latter relative to the former values is $\sim -1.8 \%$ near $t = 4 \times 10^6 \rm \, yr$. However, in our model the quantity $\dot{T}$ is of higher significance than $T$ itself; it is therefore interesting to remark that these two graphs exhibit almost equal time rates; in addition, their average time rates over the whole turn-over time are almost equal: $\left<\dot{T}_{FB}\right>_t \simeq \left<\dot{T}\right>_t = - 3.754 \times 10^{32} \rm \, erg \, s^{-1}$. \label{fig11}}

\figcaption[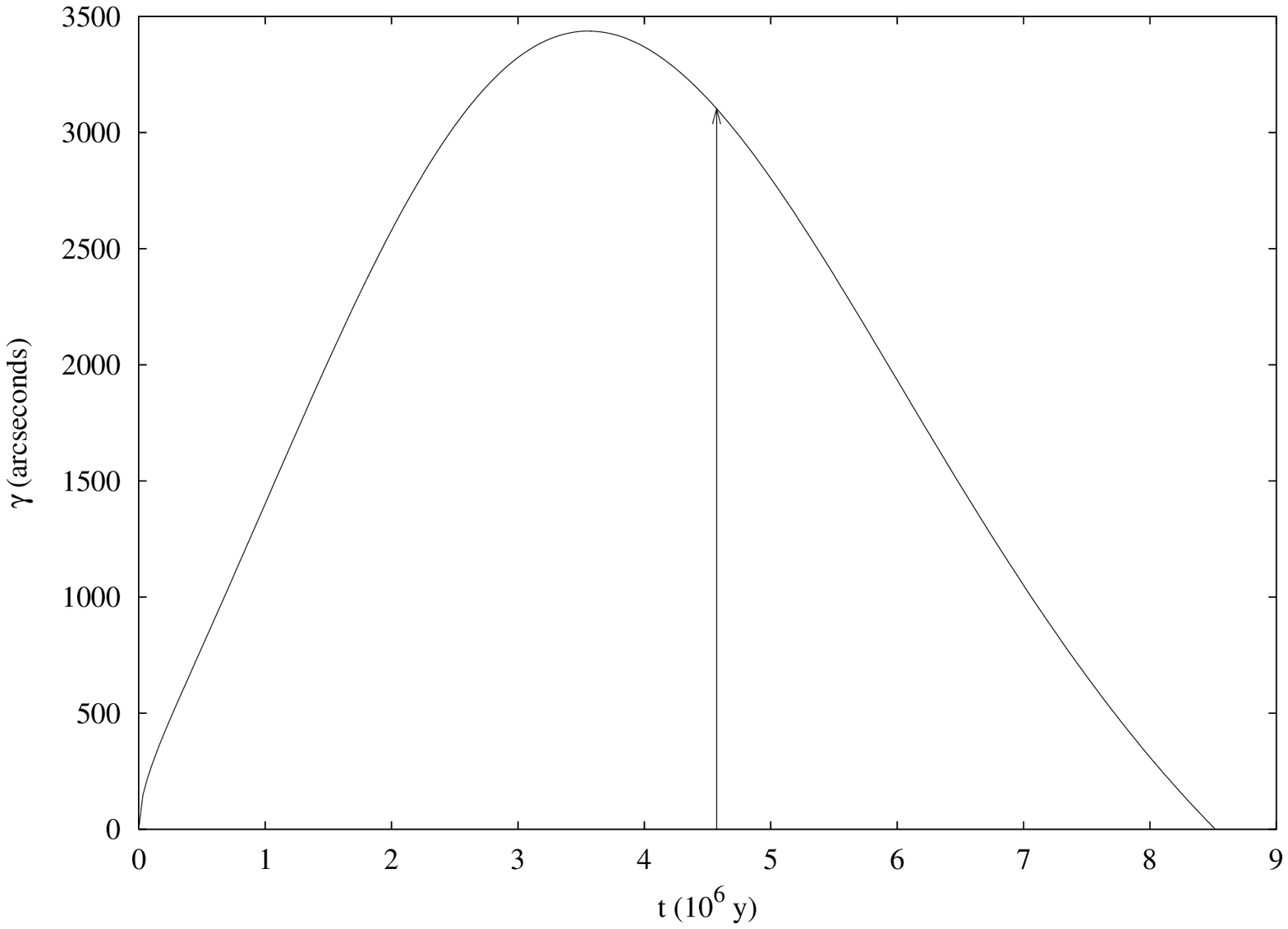]{The angle $\gamma(t)$ between the instantaneous angular velocity axis and the invariant angular momentum axis vs. $t$ in the interval $[0,t_{TOV}]$. The arrow head shows the current value $\gamma_{now} = \gamma(t_{now}) = 3102 \rm \, arcseconds$. \label{fig12}}

\clearpage
\plotone{f1.eps}
\clearpage
\plotone{f2.eps}
\clearpage
\plotone{f3.eps}
\clearpage
\plotone{f4.eps}
\clearpage
\plotone{f5.eps}
\clearpage
\plotone{f6.eps}
\clearpage
\plotone{f7.eps}
\clearpage
\plotone{f8.eps}
\clearpage
\plotone{f9.eps}
\clearpage
\plotone{f10.eps}
\clearpage
\plotone{f11.eps}
\clearpage
\plotone{f12.eps}
\clearpage

\end{document}